\documentclass[reprint,
superscriptaddress,
amsmath,amssymb,
aps,prd,floatfix,
twocolumn,
showpacs,
]{revtex4-2}

\usepackage{amsthm}
\usepackage{tikz}
\usepackage{graphicx}% Include figure files
\usepackage{dcolumn}% Align table columns on decimal point
\usepackage{bm}% bold math
\usepackage{braket}
\usepackage[english]{babel}
\usepackage{hyperref}% add hypertext capabilities
\usepackage{amsmath}
\usepackage{graphicx}
\usepackage{float}
\usepackage{siunitx}
\usepackage{xcolor}
\usepackage{xspace}
\usepackage{eucal}
\usepackage{multirow}
\usepackage[utf8]{inputenc}
\usepackage{pgfplots}
\pgfplotsset{compat=1.14}
\usepackage{upgreek}
\usepackage{siunitx}
\usepackage{gensymb}
\usepackage{amsmath}
\usepackage[justification=justified,singlelinecheck=false]{caption}
\usepackage{subcaption}
\usepackage{textcomp}
\usepackage{comment}
\usepackage{cleveref}
\usepackage[normalem]{ulem}

\Crefname{figure}{Figure}{Figures}

\usepackage{todonotes}
\usepackage{tabularx}
\usepackage{booktabs}
\usepackage{comment}
\usepackage{array}
\usepackage{url}

\begin{document}

\title{Picometer Sensitive Prototype of the Optical Truss Interferometer for LISA}

\author{Kylan Jersey}
\affiliation{Wyant College of Optical Sciences, The University of Arizona, 1630 E. University Blvd, Tucson, AZ~85719, USA}
\author{Harold Hollis}
\affiliation{Department of Physics, The University of Florida, 2001 Museum Rd, Gainesville, FL~32611, USA}
\author{Han-Yu Chia}
\affiliation{Department of Physics, The University of Florida, 2001 Museum Rd, Gainesville, FL~32611, USA}
\author{Jose Sanjuan}
\affiliation{Texas A\&M University, Aerospace Engineering, 701 H.R. Bright Bldg, College Station, TX~77843, USA}
\affiliation{Wyant College of Optical Sciences, The University of Arizona, 1630 E. University Blvd, Tucson, AZ~85719, USA}
\author{Paul Fulda}
\affiliation{Department of Physics, The University of Florida, 2001 Museum Rd, Gainesville, FL~32611, USA}
\author{Guido Mueller}
\affiliation{Max Planck Institute for Gravitational Physics (Albert Einstein Institute), Hannover, Germany}
\affiliation{Department of Physics, The University of Florida, 2001 Museum Rd, Gainesville, FL~32611, USA}
\affiliation{Leibniz University Hannover, Institute for Gravitational Physics, Hannover, Germany}
\author{Felipe Guzman}\email[Felipe Guzman; E-mail: ]{felipeguzman@arizona.edu}
\affiliation{Wyant College of Optical Sciences, The University of Arizona, 1630 E. University Blvd, Tucson, AZ~85719, USA}

%\date{\today} % Leave empty to omit a date

\begin{abstract}
The optical truss interferometer (OTI) is a contingent subsystem proposed for the LISA telescopes to aid in the verification of a $1 \frac{\mathrm{pm}}{\sqrt{\mathrm{Hz}}}$ optical path length stability. Each telescope would be equipped with three pairs of compact fiber-coupled units, each forming an optical cavity with a baseline proportional to the telescope length at different points around the aperture. Employing a Pound-Drever-Hall approach to maintain a modulated laser field on resonance with each cavity, the dimensional stability of the telescope can be measured and verified. We have designed and developed prototype OTI units to demonstrate the capability of measuring stable structures, such as the LISA telescope, with a $1 \frac{\mathrm{pm}}{\sqrt{\mathrm{Hz}}}$ sensitivity using a set of freely mountable fiber-injected cavities. Aside from its initial motivation for the telescope, the OTI can also be readily integrated with other systems to aid in ground testing experiments. In this paper, we outline our experimental setup, measurement results, and analyses of the noise limitations.
\bigbreak

\end{abstract}

\maketitle
\section{Introduction}
\label{sec:Intro}

The Laser Interferometer Space Antenna (LISA) is a planned gravitational wave detector to be operated in space to measure the gravitational wave spectrum from 0.1 mHz to 1 Hz \cite{thorpe_lisa_2010,shaddock_space-based_2008,wanner_space-based_2019}. The detector will consist of three spacecraft, each separated by about 2.5 million km, linked via laser relays to create the long baseline interferometers that form the basis of the measurement of passing gravitational waves. Each spacecraft houses free falling test masses which serve as the endpoints of the interferometers, such that the separation between two test masses aboard separate spacecraft can be monitored interferometrically. Incoming gravitational waves will create small perturbations in the separation between the test masses which LISA will measure with a precision of 10 $\frac{\mathrm{pm}}{\sqrt{\mathrm{Hz}}}$ at mHz frequencies \cite{jennrich_lisa_2009}. The LISA telescopes are used to relay both the incoming and outgoing laser beams between each spacecraft, and as such they will lie directly in the optical path of the interferometers. Fluctuations in the optical path length of each telescope will contaminate the overall measurements, leading to a $1 \frac{\mathrm{pm}}{\sqrt{\mathrm{Hz}}}$ stability requirement. If such a stability cannot be guaranteed, the length changes have to be monitored and subtracted from the LISA data streams. For this purpose, we have designed, fabricated, and tested the performance of a compact, fiber-based optical truss interferometer (OTI) that can be implemented both for ground testing and as a contingency for the flight unit telescopes.

\begin{figure}[H]
    \centering
    \includegraphics[width=\linewidth]{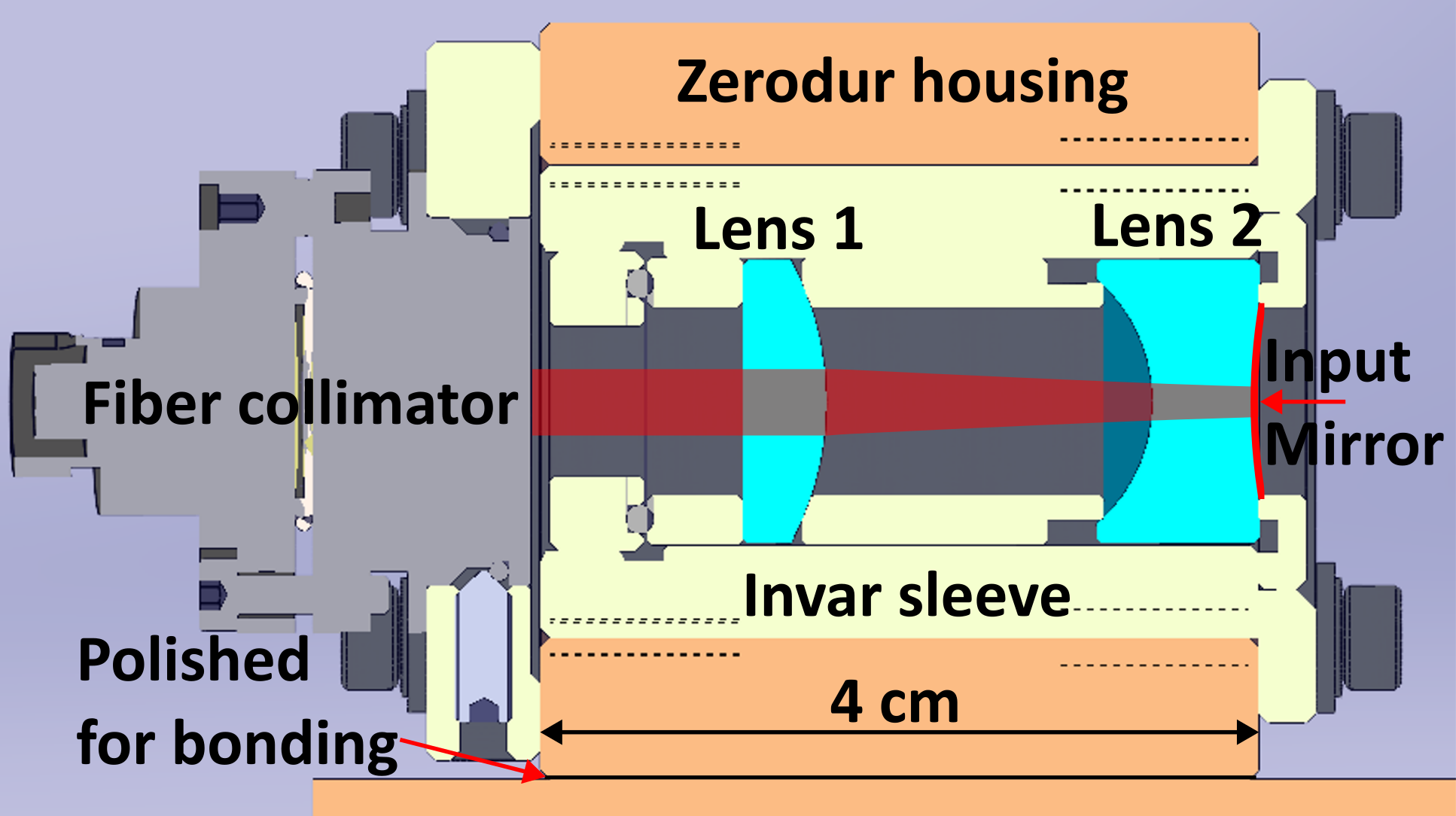}
    \caption{Layout of the optical truss input stage. An adjustable fiber collimator is secured to the housing, and two lenses are used to mode match the cavity beam. Lens 2 serves as both a mode matching lens and the cavity input mirror. Adapted with permission from Jersey et al. (2023) \textcopyright{} Optica Publishing Group \cite{jersey_optical_2023}.}
    \label{OTI design}
\end{figure}

\begin{figure*}[htpb]
    \centering
    \includegraphics[width=17.75cm]{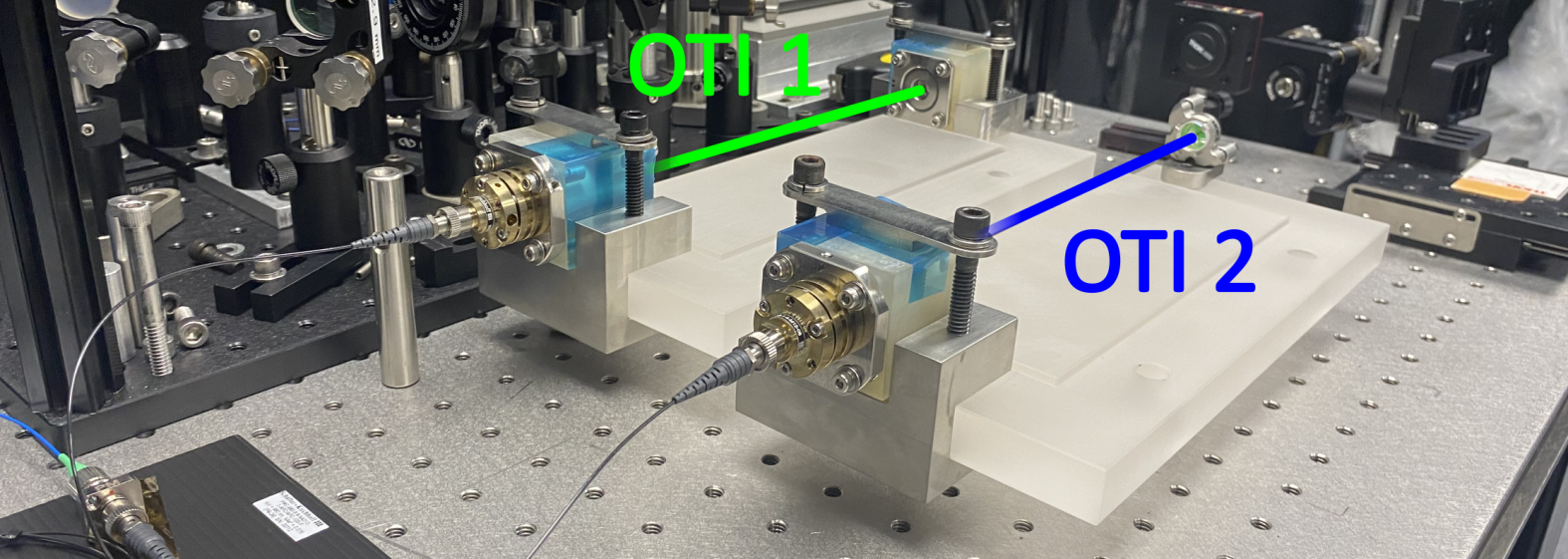}
    \caption{Photograph of the assembled and aligned prototype optical truss cavities. Each ULE plate is supported by three PEEK (polyether ether ketone) legs. OTI 1 is formed with the original input and return stages, while OTI 2 has a modified return mirror mounted to a custom post anchored in a through hole in the ULE.}
    \label{cavity pic}
\end{figure*}

The OTI is designed to serve as a ``plug and play'' solution for $1 \frac{\mathrm{pm}}{\sqrt{\mathrm{Hz}}}$ dimensional stability testing in the mHz LISA band. We have developed a compact mode matching system, or input stage, that transforms fiber-coupled 1064 nm light into the proper free-space Gaussian mode for the optical cavity within an input stage that is only 4 cm in length. While Lens 1 in Figure \ref{OTI design} serves as a positive focusing lens for mode matching, Lens 2 serves as both the second mode matching lens as well as the cavity input mirror. The front surface of the optic has a concave curvature and a negative focal length, serving as the second mode matching lens, while the backside has a concave 500 mm radius of curvature with a dielectric mirror coating with $R \geq 99.8 \%$. The input stage is accompanied with a return stage used only to house the cavity return mirror that is nominally identical to the input mirror such that they form a 70 cm long, symmetric, and impedance-matched cavity. All the internal components of the input and return units are housed in a block of Schott’s ZERODUR\textsuperscript{\textregistered{}}, polished on one side to allow for hydroxide catalysis bonding to a LISA telescope or other test structure \cite{elliffe_hydroxide-catalysis_2005,zerodur}. The optics are held in place with an Invar\textsuperscript{\textregistered{}} sleeve, and the mounting is designed to maintain the position of the cavity mirrors with respect to the zerodur housings under thermal variations. While using Invar to mount the optics is convenient for ground-based prototypes, it should be noted that the LISA spacecraft must maintain strict magnetic cleanliness and Invar would not be an acceptable choice to include in any flight hardware. Further details on the design, assembly, and alignment of these prototype units can be found in Jersey et al., 2023 \cite{jersey_optical_2023}. In principle, all that is needed to form a cavity baseline to test the dimensional stability is a suitable test structure and a mounting or bonding method to maintain the alignment between the input and return stages. In this paper, we will discuss the experimental setup and results from our initial performance testing of a first-generation prototype optical truss interferometer.

\section{Experimental Testbed}
\label{sec:experiment}
\vspace{-0.2cm}
\subsection{Optical Cavities}
\label{sec:2.1}
\vspace{-0.2cm}

We have assembled two separate OTI cavity configurations for our initial performance testing experiments. As shown in Figure \ref{cavity pic}, OTI cavity 1 is formed with the original input and return stages, and OTI cavity 2 is formed with the same input stage but with an alternative return mirror held by a commercially available opto-mechanical mount with adjustable alignment degrees of freedom. Both cavities are mounted onto plates of ultra-low expansion glass (Corning\textsuperscript{\textregistered{}} ULE\textsuperscript{\textregistered{}}) to serve as the cavity baselines \cite{ULE_glass}. Since these ULE glass plates were readily available from a previous experiment, and due to space constraints on our in-vacuum optical bench, we shortened the OTI cavity baselines from the originally designed 70 cm down to 20 cm. The input and return stages are mounted with custom stainless-steel clamps that press down on the units orthogonal to the horizontal cavity baseline. Furthermore, we used aluminum foil strips as shims to control the pitch of the input and return stages and align the cavity. Due to the shortened baseline and limited degrees of freedom, the presence of high-order transverse modes that can resonate in OTI 1 is inevitable. OTI 2, an alternative cavity configuration to compare against OTI 1, was made by procuring a suitable return mirror to account for the shortened baseline and mounting it to a ThorLabs low-distortion Polaris kinematic mount anchored to a through-hole in the ULE plate on a custom stainless-steel post, as shown in Figure \ref{cavity pic}. Cavities formed with Polaris mirror mounts similarly anchored into ULE plates have been previously shown to exhibit $1 \frac{\mathrm{pm}}{\sqrt{\mathrm{Hz}}}$ stability \cite{kulkarni_ultrastable_2020}. The alternate mirror has a convex 500 mm radius of curvature and a nominal reflectivity of $R \geq 99.8 \%$ to match the original cavity mirrors. This, along with the adjustable alignment of the Polaris mount, drastically reduced the presence of high-order modes in the OTI 2 cavity. These two trial cavity configurations exemplify the ``plug and play'' nature of the OTI, as the cavity baseline can be adjusted for any test structure given a proper design of the return mirror and mounting, and the input stages can always accommodate the cavity injection.

\begin{figure*}[htpb]
    \centering
    \includegraphics[width=13cm]{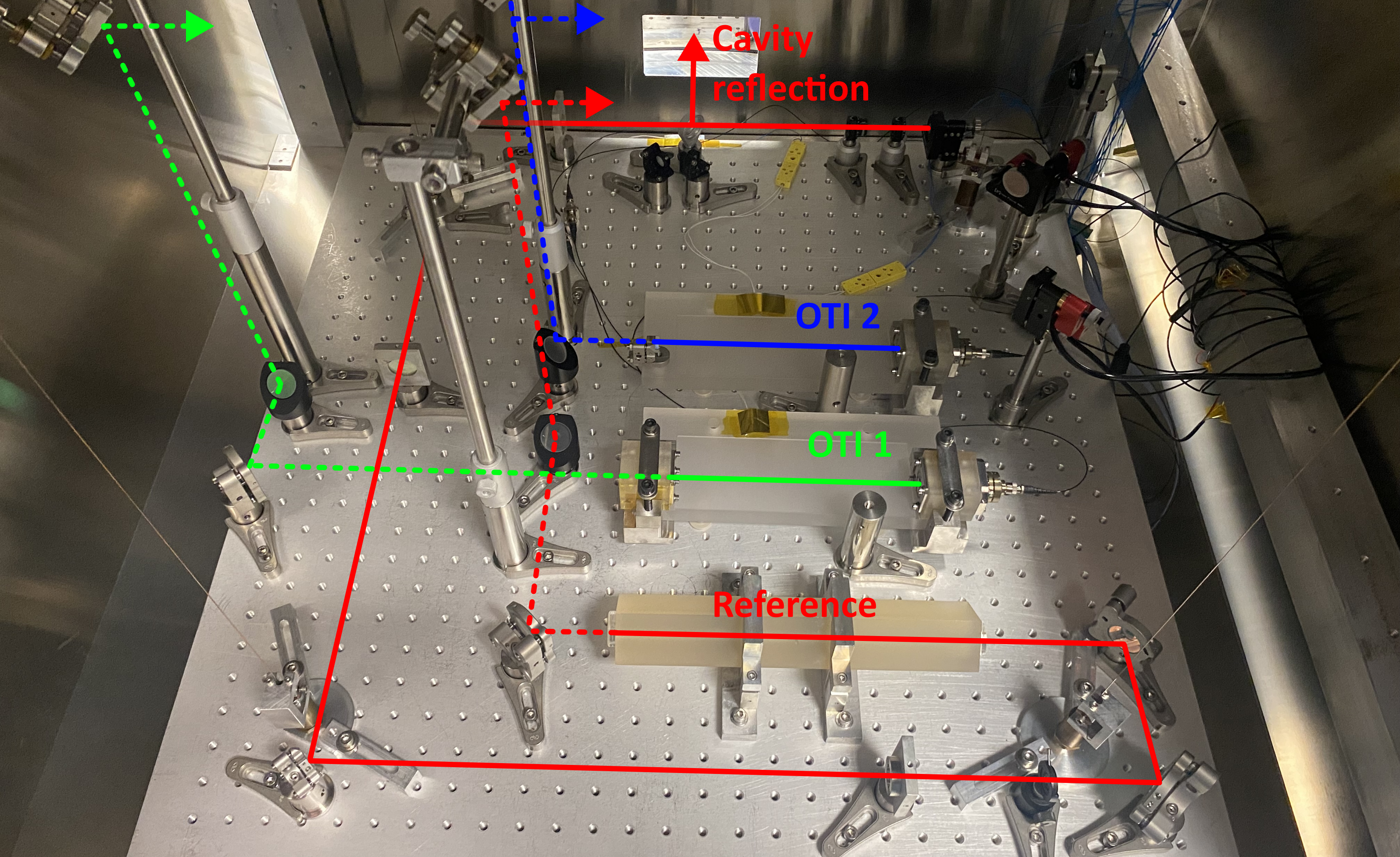}
    \caption{Photo of the in-vacuum optical bench with the optical paths for each cavity labeled. The reference cavity input and reflection is coupled in free space, while the OTI cavities are fiber-injected. The cavity transmission beams (dashed lines) are routed out of the chamber through a view-port window via periscopes.}
    \label{in vac bench}
\end{figure*}

Our reference cavity is a zerodur spacer with optically-contacted, high-finesse cavity mirrors bonded on either end of the spacer. This cavity has been verified with similarly fashioned reference cavities to have a stability of 100 $\frac{\mathrm{Hz}}{\sqrt{\mathrm{Hz}}}$ or better in the LISA band, which serves as a sufficient laser frequency standard for the heterodyne readout of the OTI cavities \cite{kulkarni_technology_2022}. The injection into the reference cavity is prepared in free space on the in-vacuum optical bench after passing through a fiber-coupled feed-through flange into the vacuum chamber. The cavity reflection is isolated via a polarizing beam-splitter (PBS) and quarter-wave plate (QWP) and is sent out of the vacuum chamber through a view-port window to be measured by a photodetector (PD), as seen in Figure \ref{in vac bench}. The cavity transmission, as with the transmission through the OTI cavities, is sent out of another view-port window and measured as a diagnostic for the PDH locking. We characterize the contrast in the resonance of the fundamental Gaussian mode, $V_{00}$, by measuring the ratio of the reflected power that vanishes on resonance to the total reflected power away from resonance. We can also inspect the high-order mode content of each cavity by monitoring the cavity transmission with a CMOS camera for each resonance that is observed. The relevant optical parameters of each cavity, such as the free spectral range (FSR), full-width half-maximum (FWHM) linewidth, and $V_{00}$, are shown in Table \ref{table:cavities}.

\begin{table}[htpb]
    \centering
\begin{tabular}{|c||c|c|c|}
\hline
& \textbf{FSR} & \textbf{Linewidth} &  \\

  \textbf{Cavity} & \textbf{(Cavity Length)} & \textbf{(Finesse)} & \textbf{\emph{V}\textsubscript{00}} \\
\hline \hline
    &  FSR = 758 MHz & $\Delta\nu = 210 kHz$ & \\
 OTI Cavity 1 & ($L$ = 19.8 cm) & $(\mathcal{F} = 3610)$ & 74\% \\
\hline
  &  FSR = 726 MHz & $\Delta\nu = 300 kHz$ & \\
 OTI Cavity 2 & ($L$ = 20.7 cm) & $(\mathcal{F} = 2420)$ & 71\% \\
\hline
 & FSR = 578 MHz & $\Delta\nu = 100 kHz$ & \\
 Reference Cavity & ($L$ = 25.9 cm) & $(\mathcal{F} = 5780)$ & 82\% \\
\hline
\end{tabular}
\caption{Optical properties of the cavities used for the OTI experiments. These measurements were made by scanning the laser frequency across each cavity resonance and calculating the various parameters from the response of the cavity reflections.}
\label{table:cavities}
\end{table}

\subsection{Vacuum Chamber Isolation}
\label{sec:2.2}

\begin{figure*}[htpb]
    \centering
    \includegraphics[width=14cm]{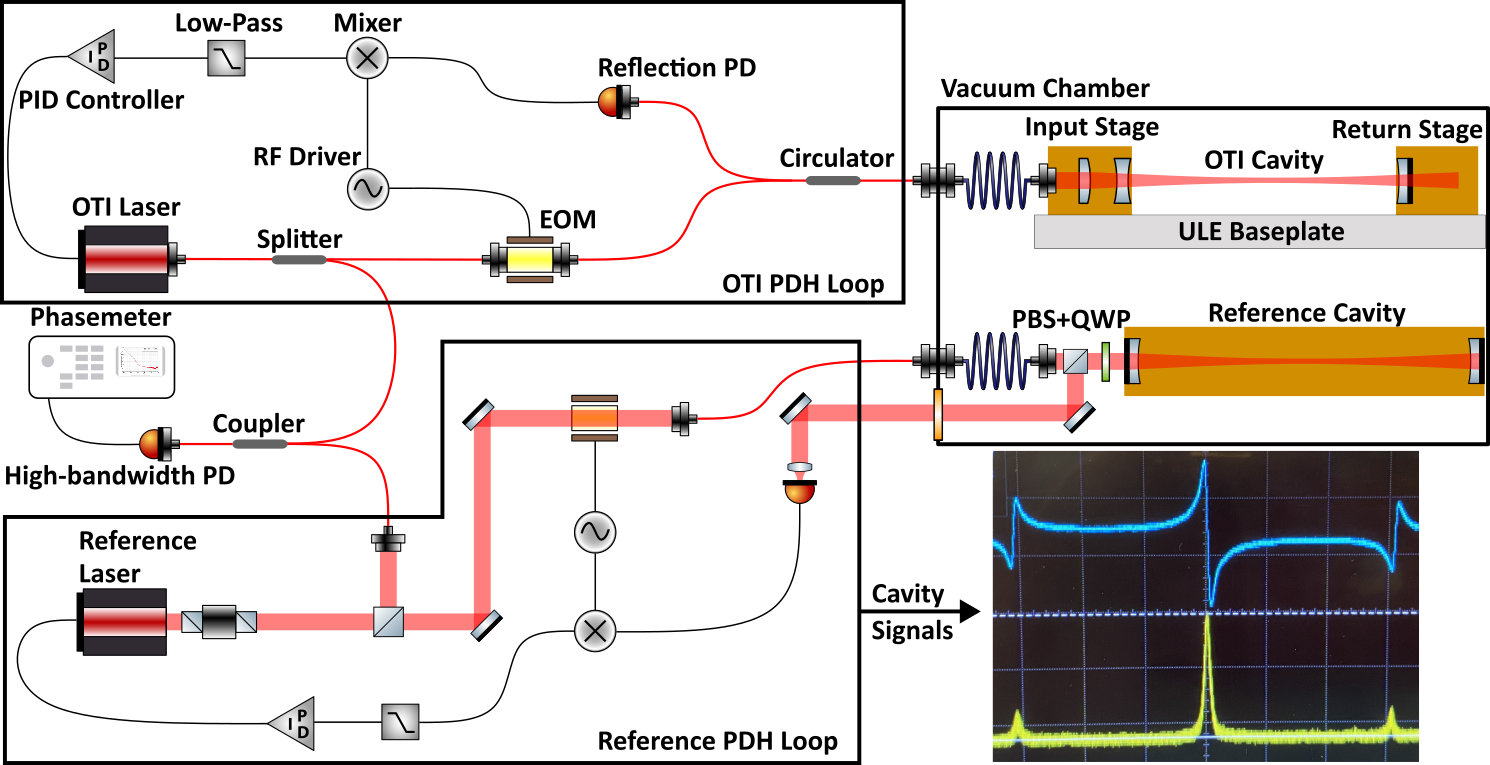}
    \caption{Cavity readout scheme for the OTI experiment. The OTI laser path, including the cavity injection and reflection, is purely fiber-coupled while the reference laser is primarily coupled in free space apart from the fiber feed-through into the vacuum chamber. The figure shows an example of the reflected power and resulting PDH error signal as the laser frequency is scanned over a resonance. The two laser fields are combined in a fiber coupler to interfere on a high-bandwidth PD.}
    \label{PDH setup}
\end{figure*}

\begin{figure}[h]
    \centering
    \includegraphics[width=\linewidth]{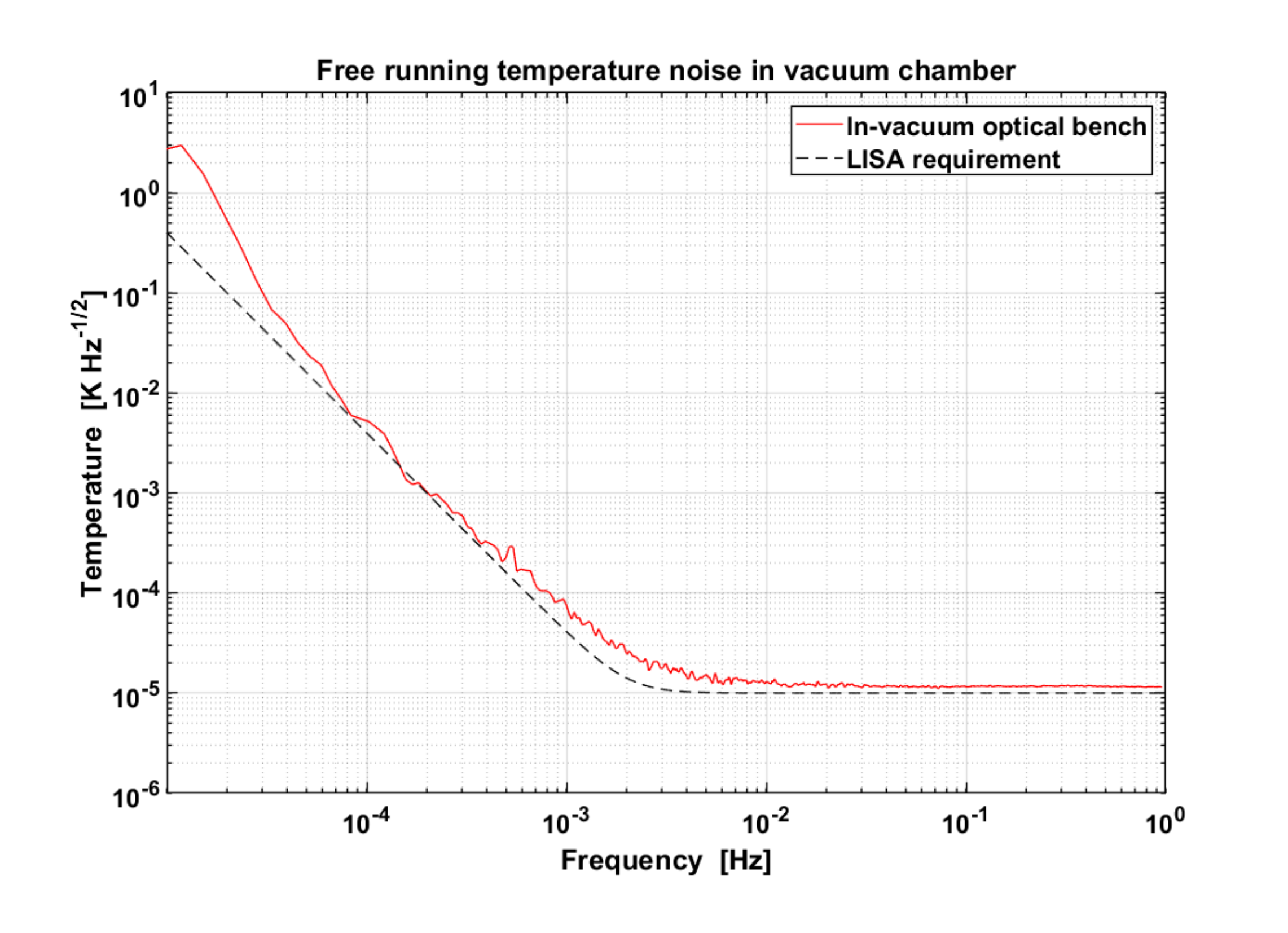}
    \caption{Temperature stability as measured by thermistors attached to the cavities and optical bench in our vacuum chamber, and compared with the benchmark LISA temperature requirement.}
    \label{temp stability}
\end{figure}

The optical cavities must be operated in a thermally shielded and mechanically isolated vacuum chamber to simulate the environment that the LISA spacecraft are likely to create \cite{peabody_low_2006,sanjuan_mathematical_2015}. To achieve this, we utilize a large steel chamber, suspended on coil springs, which houses an optical bench suspended by four phosphor-bronze wires and surrounded by aluminum thermal shield walls on all sides. The suspension wires hang from blade springs attached to a large aluminum frame that is contacted with the bottom of the chamber only through MACOR\textsuperscript{\textregistered{}} spacers, providing low thermal conduction from the chamber exterior \cite{macor}. We can measure the temperature stability inside the chamber via thermistors whose resistance will change with temperature according to the Steinhart-Hart equation with coefficients which can either be measured or provided by the manufacturer \cite{steinhart_calibration_1968}. The resistance of each thermistor is recorded as a time series, and the amplitude spectral density of the temperature noise can be calculated and compared with the benchmark LISA requirements, which is shown in Figure \ref{temp stability}.

The free running temperature noise that we measure in the vacuum chamber is typically on par with the LISA requirements above 0.1 mHz. The flattening of the spectrum seen around 1 mHz is a readout limitation due to the analog-to-digital conversion (ADC) used by the multimeter, and the true temperature noise at and above 1 mHz is estimated to be well below the requirement \cite{kulkarni_technology_2022}. The noise below 0.1 mHz is above the LISA requirement likely because we allow the temperature to freely run rather than using active stabilization. Even without active temperature control, the cavity spacers and plates we are using, made of zerodur and ULE which have CTE on the order of $10-100\times10^{-9}$\,K$^{-1}$, are stable enough to still meet the $1 \frac{\mathrm{pm}}{\sqrt{\mathrm{Hz}}}$ LISA requirement in this environment. However, it is the OTI input and return stages, as well as the steel mounts holding them to the plates, that may dominate the thermal expansion of the cavities. Thus, it is the effective CTE to which each individual component contributes that will limit the cavity displacement noise at low frequencies where temperature noise dominates \cite{sanjuan_long-term_2019,numata_thermal-noise_2004,kessler_thermal_2012}.

\subsection{Cavity Readout Scheme}
\label{sec:2.3}

The foundation of how we measure the displacement noise in the optical truss cavities lies in the Pound-Drever-Hall (PDH) frequency locking method. PDH locking involves creating a feedback loop that maintains the laser frequency on resonance with an optical cavity. By phase modulating the incident field and measuring the field reflected from the cavity, an error signal can be generated, filtered, and fed back into the laser controller to maintain resonance. For the OTI, we employ commercially available nonplanar ring oscillator (NPRO) lasers and fiber-coupled electro-optic phase modulators (EOMs) that are placed in the optical path before coupling into polarization-maintaining optical fibers that feed into the vacuum chamber and into each respective cavity. The reflected field from each cavity is isolated from the incident field via an optical circulator. This is done with a PBS and QWP for the reference cavity, and with commercially available fiber-optical circulators for the OTI cavities. By measuring the reflected power with a photodetector and subsequently demodulating with the RF signal driving the phase modulation, the result is an error signal with a zero-crossing that is linearly proportional to the deviation from resonance \cite{black_introduction_2001}. The PDH error signal is filtered through proportional-integral (PI) servos and fed back into the laser controllers, which stabilize the laser frequency to the cavity resonance. While the laser is locked to an optical cavity, the laser frequency is coupled with the optical path length fluctuations by the relation

\begin{equation}
    \frac{\delta \nu}{\nu} = -\frac{\delta L}{L}
    \label{eq1}
\end{equation}

\noindent where $\nu$ is the laser frequency and $L$ is the cavity length, and $\delta \nu$ and $\delta L$ are their fluctuations, respectively. We measure the stabilized laser frequency noise, $\delta f$, by forming an RF beat frequency between a laser locked to an OTI cavity and another laser locked to our reference cavity, as depicted in Figure \ref{PDH setup}. In each laser’s optical path, before the EOM phase modulation, there is a pick-off beam created with a beamsplitter to use for beat signal generation. The pick-off beams from each laser are combined on a high-bandwidth PD to produce an RF beat frequency equal to the difference between the optical frequencies of the two lasers. The primary measurement is made using a digital FPGA-based Moku:Lab phasemeter to record the beat frequency between two lasers while one is locked to an OTI cavity and the other to the reference cavity \cite{moku_phasemeter}.

We employ a spectral analysis approach to estimate the amplitude spectral density (ASD) of the beat frequency time series, which can then be converted to the corresponding cavity length noise via Equation \ref{eq1} assuming that the variations in the beat frequency are dominated by the OTI laser and not the reference \cite{stoica_spectral_2005,welch_use_1967}. We employ the LTPDA signal processing toolbox in MATLAB\textsuperscript{\textregistered{}}, not only to calculate the ASD of time series data but also to perform many other reliable data analysis tasks in both the frequency and time domain \cite{hewitson_data_2009}. An example of the estimated ASD of the cavity length noise compared to the benchmark LISA requirement is shown in Figure \ref{Noise budgets} along with other estimated contributions from the various noise sources that were investigated.

\section{Results \& Analyses}
\label{sec:noise budget}

\begin{figure*}[htpb]
    \centering
    \includegraphics[width=8.75cm]{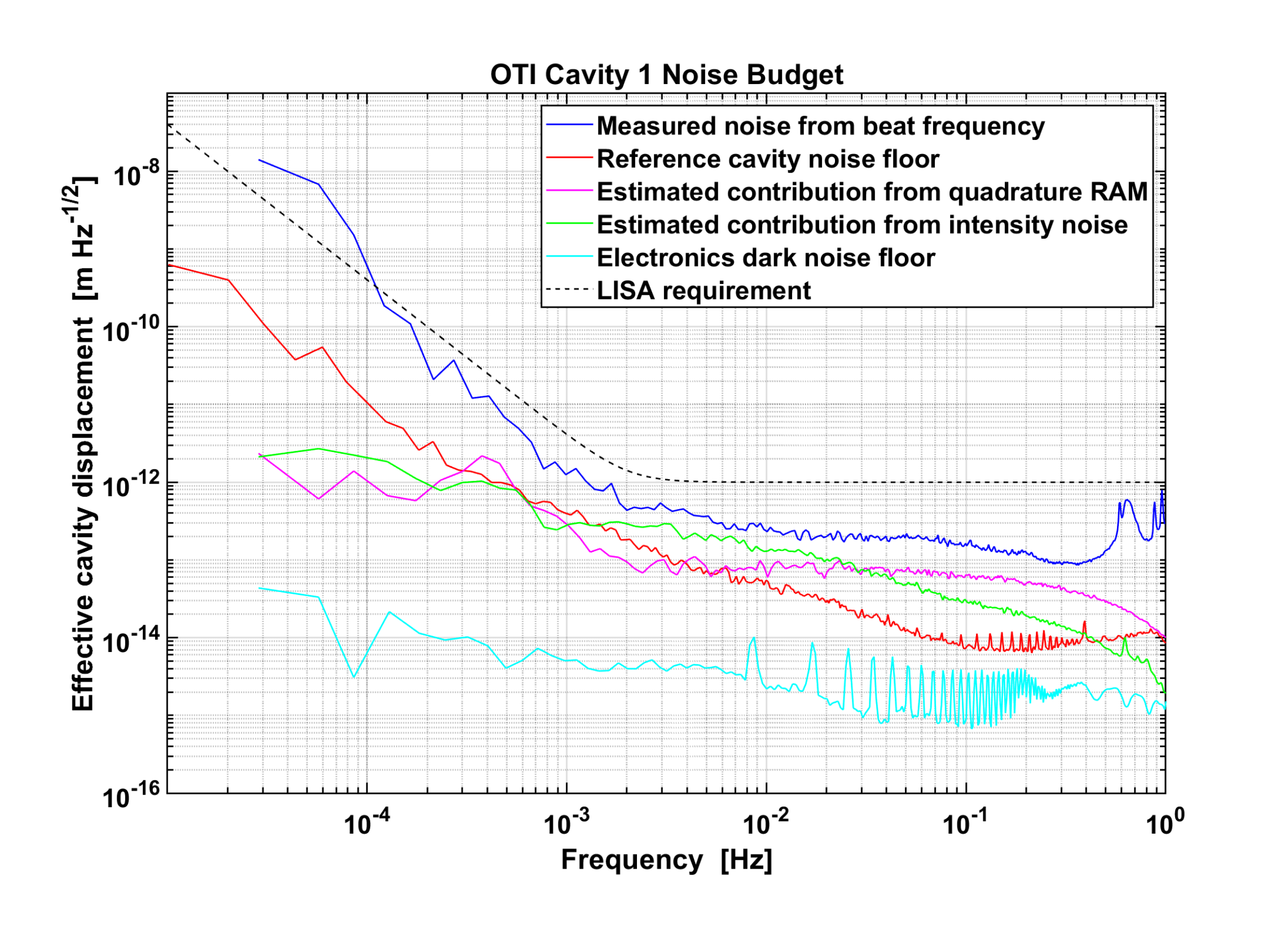}
    \includegraphics[width=8.75cm]{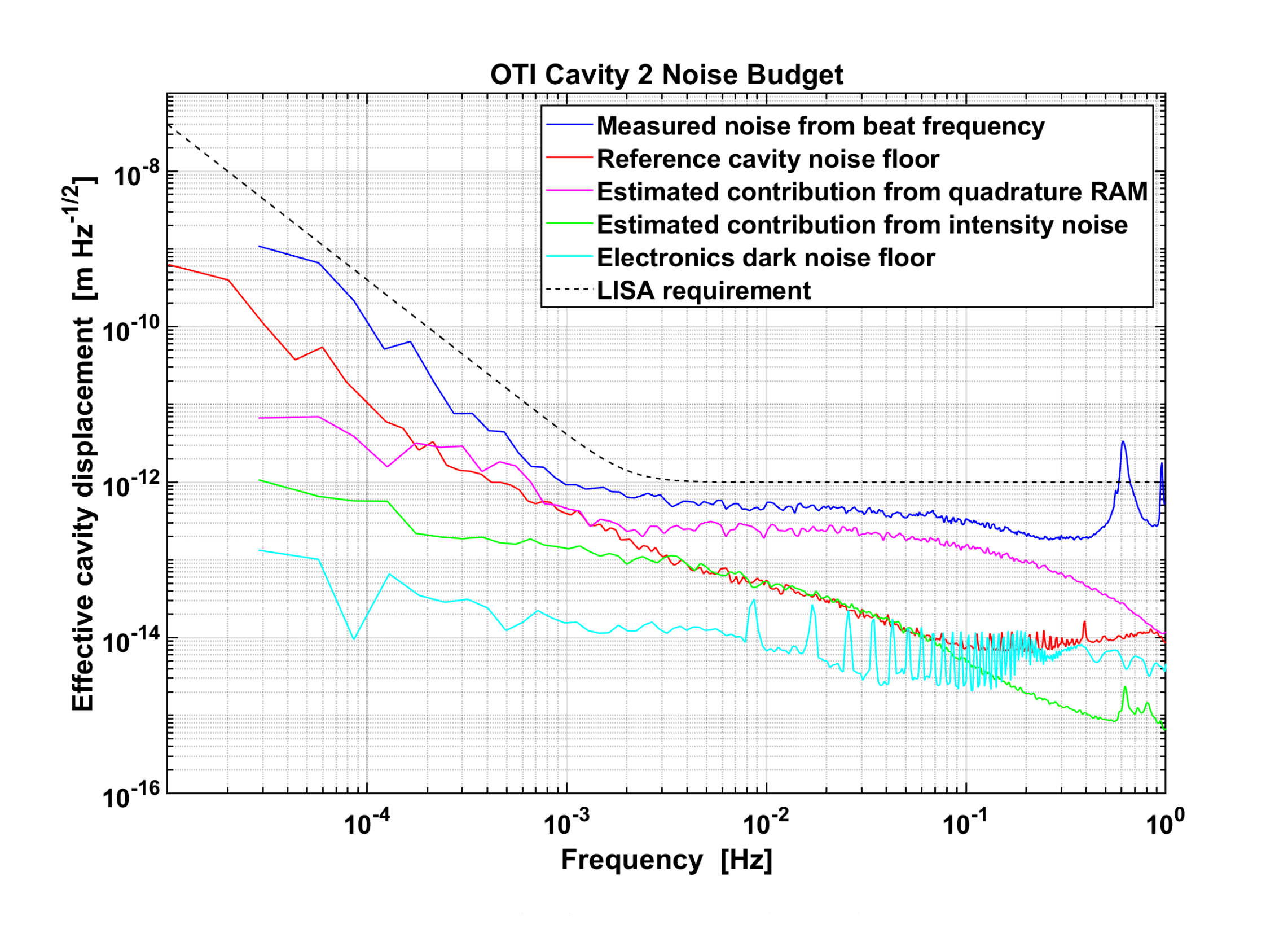}
    \caption{Noise apportioning for OTI 1 (left) and OTI 2 (right). The estimated contributions from the RAM and intensity noises were taken concurrently with the respective cavity measurements, while the contributions from the reference cavity and electronics dark noise were taken separately. The peaks around 0.6 - 1 Hz in the cavity noise spectra (blue) are from known mechanical resonances in the vacuum chamber suspension.}
    \label{Noise budgets}
\end{figure*}

The measurement results shown in Figure \ref{Noise budgets} demonstrate that we have met the LISA telescope $1 \frac{\mathrm{pm}}{\sqrt{\mathrm{Hz}}}$ sensitivity requirement between 0.1 mHz - 1 Hz with both prototype OTI test cavities. The simple construction of steel clamps pressing the input and return stages to the ULE plates, the commercial Polaris mirror mount holding the alternative cavity mirror in OTI 2, and the purely fiber-coupled cavity inputs all exemplify this powerful ``plug and play'' solution for possible ground testing experiments. The cavity measurements have unique differences at separate regions across the LISA frequency band. OTI 1 performs very well above 1 mHz, reaching as low as $0.1 \frac{\mathrm{pm}}{\sqrt{\mathrm{Hz}}}$, while OTI 2 generally exhibits a higher noise above 1 mHz and even exceeds the LISA requirement at the peaks near 1 Hz. However, OTI 2 outperforms OTI 1 below 1 mHz and is much less prone to drift, even meeting the LISA requirement below 0.1 mHz. Our investigations into the various sources that contribute to the overall measured cavity noise, as well as the differences measured between each OTI cavity, are detailed in the following sections.

\subsection{Temperature Noise}
\label{sec:3.1}

\begin{figure}[h]
    \centering
    \includegraphics[width=\linewidth]{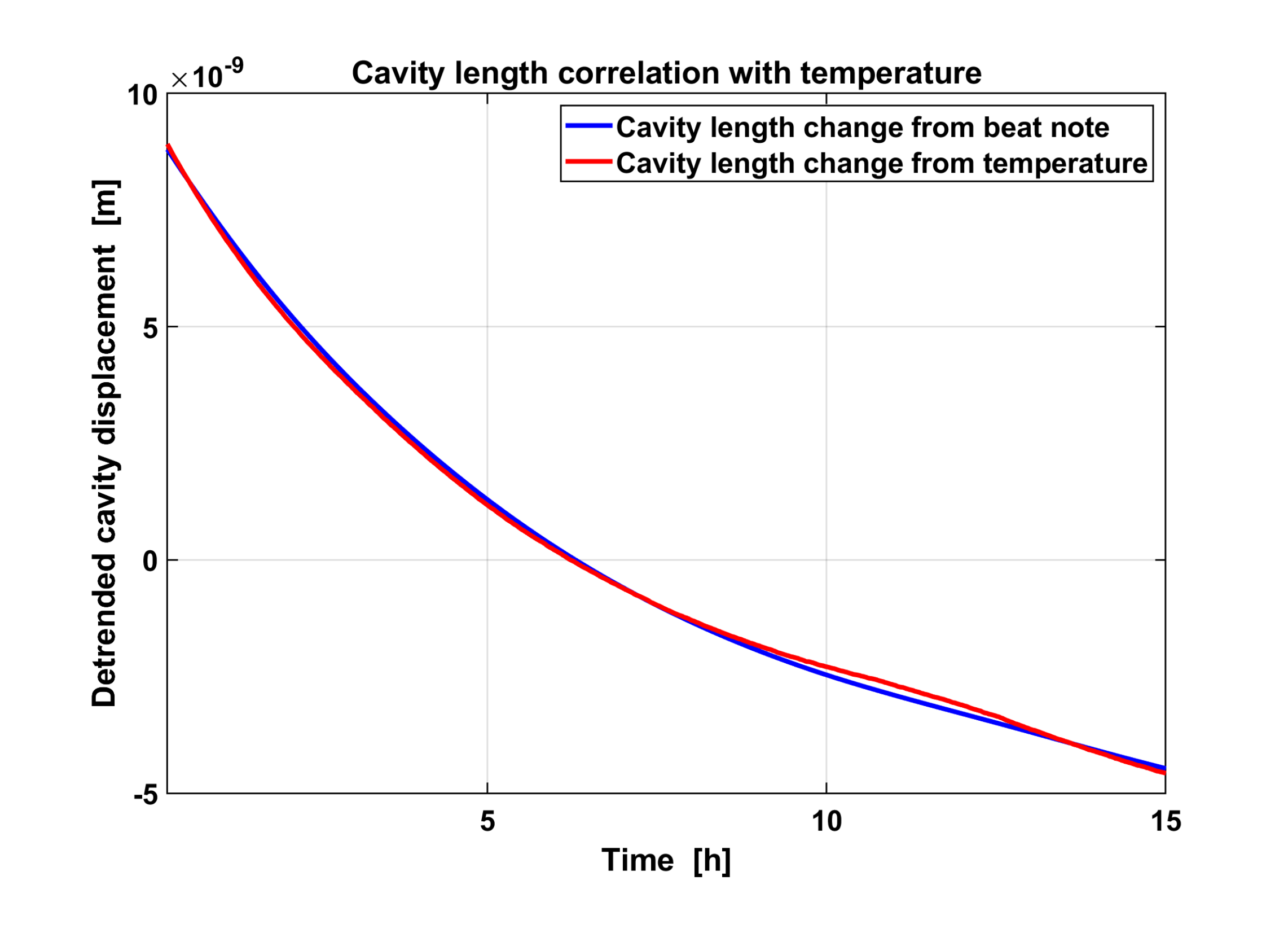}
    \caption{Plot showing the low-frequency correlation between temperature and cavity length in OTI cavity 1. The temperature is calibrated to cavity length via the CTE calculated from a least-squares fit between the two time series.}
    \label{CTE plot}
\end{figure}

Each noise source can couple into the measured cavity length noise through different mechanisms. An intuitive example is the temperature noise in the vacuum chamber coupling into the cavity length through the effects of thermal expansion. In this case, the coupling factor is the effective CTE of the optical cavity being tested, which can vary depending on the materials used in the cavity assembly, the type of cavity mirrors used, as well as the way the mirrors are bonded or mounted to the cavity spacer \cite{numata_thermal-noise_2004,kessler_thermal_2012,sanjuan_long-term_2019}. The effective CTE can be measured by taking concurrent measurements of the cavity temperature along with the primary beat note measurement and calculating the coupling factor between the two time series with a least-squares fit, depicted in Figure \ref{CTE plot}. Both time series are low-pass filtered with a 1 mHz corner frequency, below which the correlation between the temperature and beat note is dominant. Note that in our experiment, we have not actively driven the temperature in the chamber and we use the gradual drift around room temperature over many hours or days to correlate with the beat note measurements. Ideally, we would seek a frequency reference that is external to the vacuum chamber or that is temperature invariant such as a molecular transition, and actively drive the temperature to correlate with the beat note. However, we estimate a room temperature ($\approx$ 20$^{\rm o}$C in our laboratory) CTE of $-1.25 \pm 0.23\times10^{-6}$\,K$^{-1}$ ($250 \pm 50$\,nm/K) in OTI 1 and $2.0 \pm 0.4\times10^{-7}$\,K$^{-1}$ ($40 \pm 10$\,nm/K) in OTI 2, which are both much larger than what we would expect if the dominant contributions were from the ULE spacers and the zerodur blocks housing the OTI units ($\sim10^{-8}$\,K$^{-1}$). Since the effective CTE estimates are well above the CTE of zerodur or ULE-based cavities, it is a safe assumption that our measured values roughly estimate the effective CTE of the OTI cavities and that they are likely limited by the metal components and mounts in the assemblies.

The OTI input and return stages and the overall cavity assemblies are constructed with multiple different types of material such as zerodur, ULE, fused-silica, Invar, and stainless-steel. The way these various components are assembled to form the cavities leads to their individual contributions to the overall effective CTE. The original OTI input and return stages were designed such that the thermal expansion of the Invar components minimally couple into the position of the cavity mirror surfaces with respect to the zerodur housing \cite{jersey_optical_2023}. The return mirror mount for OTI 2 is similarly fashioned such that the cavity mirror is ideally positioned at the center of thermal expansion as it is anchored to the through hole in the ULE plate \cite{kulkarni_ultrastable_2020}. However, this critically depends on how well-centered the mirror surface is above the zero point of thermal expansion. Furthermore, the contribution from the steel clamps pressing down on the OTI units is also not well characterized and beyond the scope of our investigations. Overall, these minimalist mounting solutions used for the OTI cavities required no bonding and the steel parts were made in-house, yet our measurements still meet the LISA requirement at low frequencies down to 0.1 mHz. Future experiments may use different mounting or bonding methods for the OTI units to further improve the low-frequency noise and reveal more information about the input and return stages themselves.

\begin{figure*}[htpb]
    \centering
    \includegraphics[width=14cm]{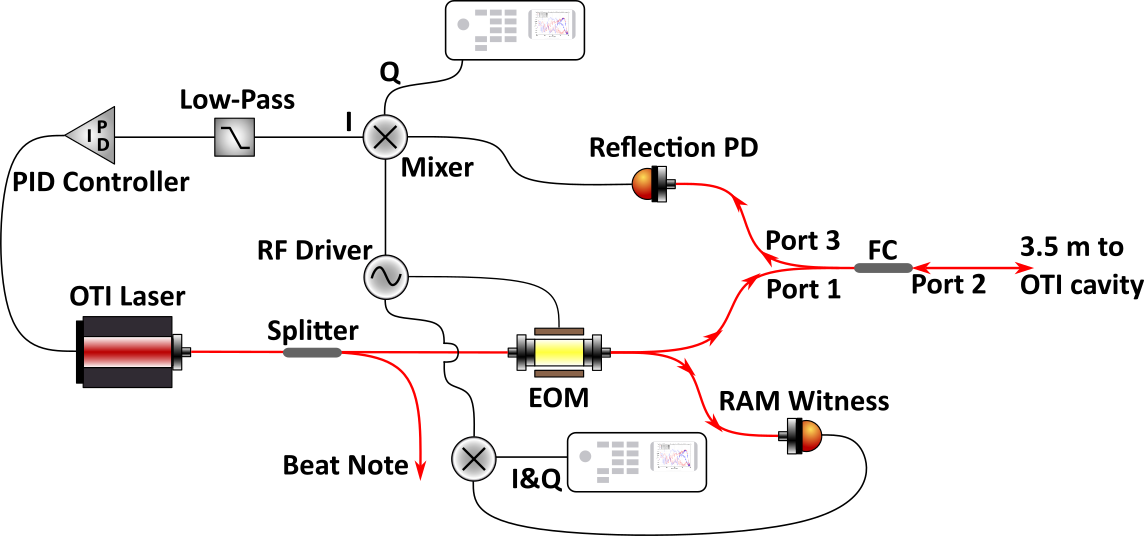}
    \caption{Configuration for measuring the RAM noise both directly after the EOM and in the cavity reflection. The parasitic interference arises in the fiber circulator (FC) between a small leakage field that couples from port 1 to port 3 and the primary cavity reflection that couples from port 2 to port 3. This interference effect dominated our initial OTI noise measurements.}
    \label{RAM setup}
\end{figure*}

\subsection{Intensity Noise}
\label{sec:3.2}

Intensity fluctuations in the cavity field can couple into the effective cavity length through thermo-elastic deformation of the cavity mirror surfaces, thermo-refractive effects in the mirror coatings, along with other effects \cite{gretarsson_three_2018,evans_thermo-optic_2008,hello_analytical_1990}. The combination of these effects will lead to an overall coupling factor between the intensity noise and the cavity length changes. To estimate this, we added a commercially available fiber-coupled electronic variable optical attenuator (VOA) into the fiber path just before the fiber EOM and circulator. The voltage applied to the VOA attenuates the output power from 0-5 V, where 0 V is maximum transmission and 5 V is maximum attenuation. While the laser is locked to one of the OTI cavities, we track the beat note between the measurement and reference lasers as well as the optical power transmitted through the OTI cavity. We then modulate the power transmitted through the VOA, and we correlate the power variations seen in the cavity transmission with the beat frequency to get a coupling coefficient. While this coefficient is generally frequency-dependent, we modulate the power slowly (10 mHz) to calculate the steady-state coupling between intensity and cavity length, which is expected to represent the noise contribution in the LISA band. We can then take the typical intensity noise seen while the laser is locked with no intentional power modulation, and project the noise contribution to the cavity measurements as shown in Figure \ref{Noise budgets}. We have measured coupling coefficients of $193 \pm 10$ MHz/W (140 nm/W) in OTI cavity 1 and $54 \pm 4$ MHz/W (35 nm/W) in OTI cavity 2, which generally are in good agreement with theoretical expectations as well as prior similar experiments \cite{sanjuan_long-term_2019,kulkarni_technology_2022,hello_analytical_1990}. Future investigations may seek to adjust the cavity parameters or to use active stabilization of the optical power in the cavity to further characterize and improve the measurements if necessary.

\subsection{Residual Amplitude Modulation}
\label{sec:3.3}

\begin{figure*}[htpb]
    \centering
    \includegraphics[width=8.75cm]{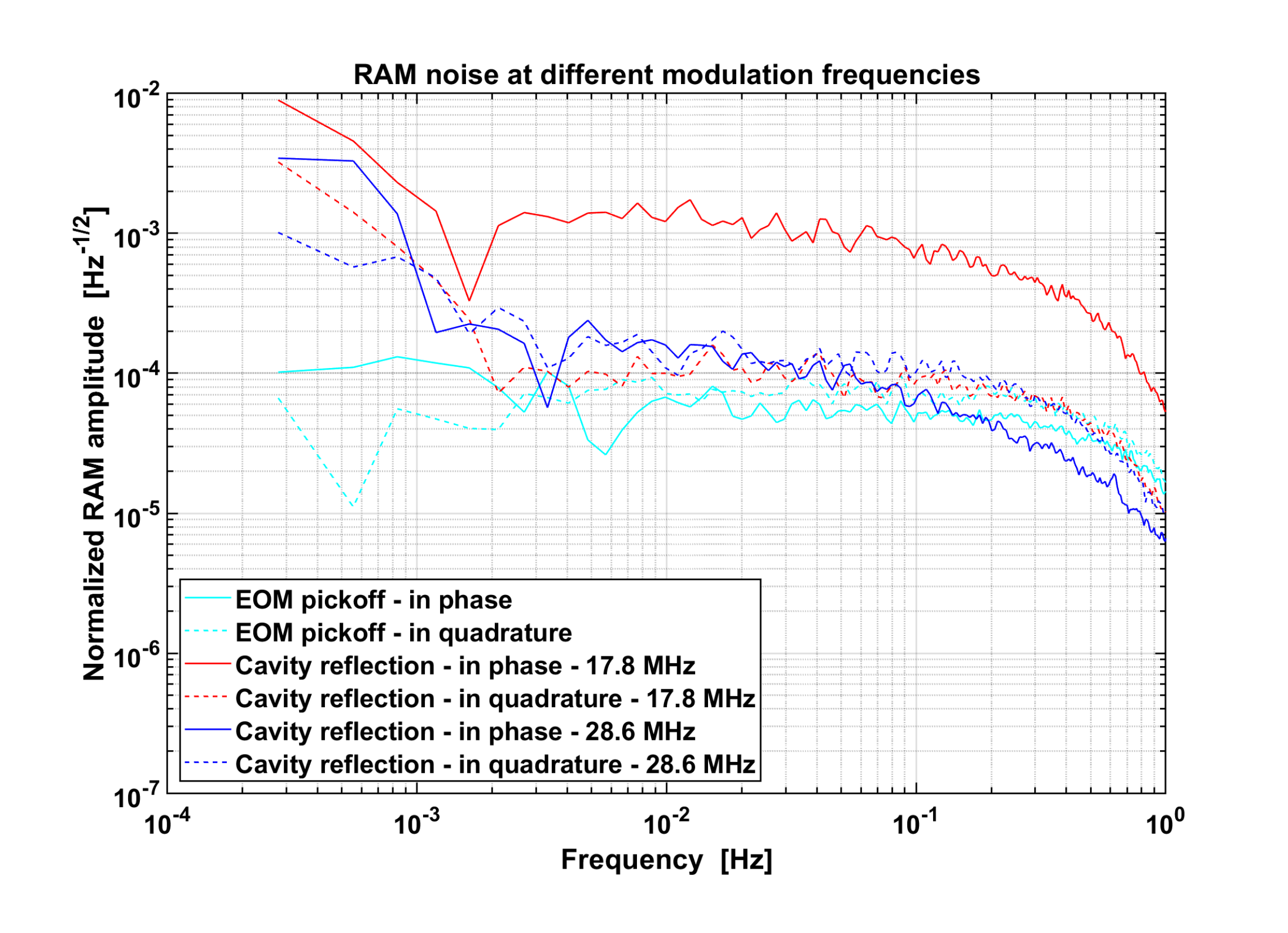}
    \includegraphics[width=8.75cm]{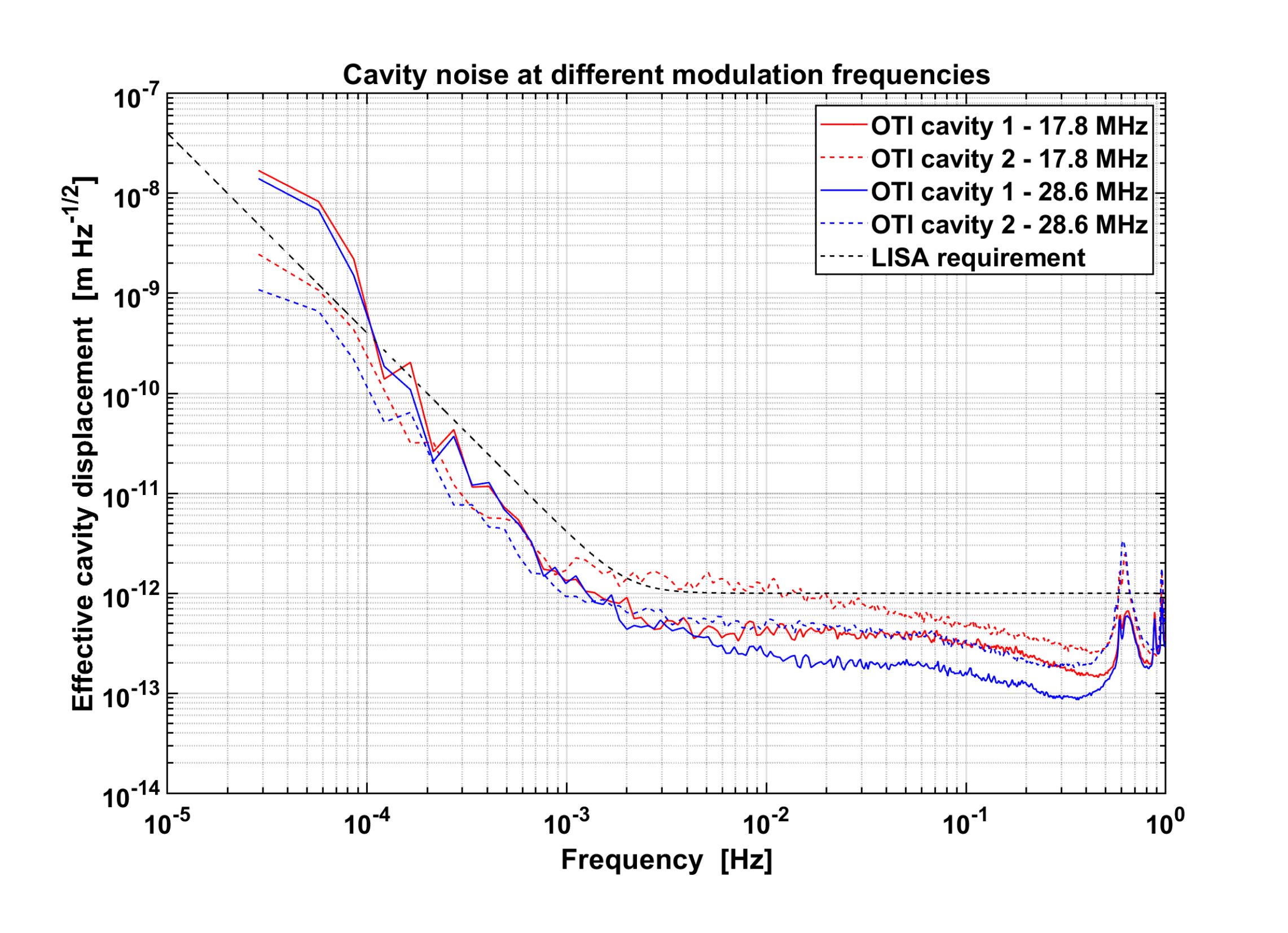}
    \caption{Comparison of the RAM noise measured in the cavity reflection at different modulation frequencies. Each term is normalized by the average optical power. The effect of changing the modulation frequency to the free spectral range of the parasitic etalon can be seen in both the RAM noise and the measured cavity noise. The red curves are from measurements with a modulation frequency of 17.8 MHz and the blue curves are with a frequency of 28.6 MHz.}
    \label{RAM plots}
\end{figure*}

Residual amplitude modulation (RAM) is a well-known noise source in PDH frequency stabilization experiments, and has been described extensively in the literature \cite{zhang_reduction_2014,wong_servo_1985,duong_suppression_2018,shi_suppression_2018,shen_systematic_2015}. A phase-modulated beam reflected from a cavity will have carrier and side-band components that interfere with each other when the laser is near resonance, and this interference is measured with a PD and demodulated to produce an error signal \cite{black_introduction_2001}. In practice, the phase modulation is not perfect and the beam transmitted through the EOM will contain a residual amplitude modulation, or RAM. There are many possible sources of RAM, including the temperature-dependent birefringence of the crystal, parasitic etalon effects from the end faces of the crystal or other reflective surfaces, the polarization extinction ratio, and the alignment of the field polarization to the ordinary and extraordinary axes of the modulator crystal. The cumulative effects of RAM are detected in the demodulation of the cavity reflection, and manifested as a distorted error signal whose zero-crossing is offset from the cavity resonance \cite{wong_servo_1985}. The effects of RAM are generally not constant in time and will translate into an apparent frequency drift around the true cavity resonance while the laser is locked.

We found that the dominant noise term above 1 mHz in our initial cavity measurements was caused by a parasitic etalon interference in the fiber path between the leakage field in the fiber-coupled circulator and the field reflected from the cavity. This parasitic interference exacerbated the RAM effect that was measured in the cavity reflection and caused fluctuations in the PDH loop and consequently in the laser frequency. We characterized this noise by monitoring the demodulated signals from the cavity reflection PD, both while the laser was locked and while the laser was parked off-resonance. However, one component of the demodulation is the PDH error signal near resonance, which we call the in-phase component for our purposes. Both the in-phase and quadrature (I and Q) RAM signals can be monitored while the laser is parked off-resonance from the cavity, but the in-phase component is suppressed by the closed PDH loop while the laser is locked. We compared the noise in the RAM measured directly after the EOM to the noise measured in the cavity reflection off-resonance, normalizing the signals to account for losses, and found that the in-phase RAM noise in the cavity reflection dominated over the other signals (Figure \ref{RAM plots}). Parasitic interference in the reflection port (port 3) of the fiber circulator, arising from a small leakage field that couples from the circulator input (port 1 $\rightarrow$ port 3) and interferes with the primary cavity reflection (port 2 $\rightarrow$ port 3) as depicted in Figure \ref{RAM setup}, has been noted in the literature in similar cavity stabilization experiments \cite{wegehaupt_optical_2024}.

The effect of parasitic interference coupling into PDH locking experiments is well known and this issue is typically avoidable \cite{shen_systematic_2015}. In free-space coupled systems, the optical elements whose surface reflections form the interference fringes can be slightly tilted to prevent the buildup of standing waves. The optical truss system, however, is meant to be entirely fiber-coupled and we cannot tilt any elements in the optical path to de-couple parasitic reflections. In previous studies, choosing the modulation frequency to match the free spectral range (FSR) of an etalon in the optical path has helped to minimize the RAM \cite{whittaker_residual_1985}. We tested different modulation frequencies between our original choice of 17.8\,MHz up to 40\,MHz, monitored the noise in the mHz band at each frequency, and found that the in-phase RAM noise reached a minimum when modulating around 30 MHz. Considering the optical path of the etalon, there is 3.5\,m of fiber between the circulator and the OTI cavity. The FSR of such an etalon formed between the circulator and the cavity, assuming the fiber index $n \approx 1.5$ and using $\mathrm{FSR} = \frac{c}{2nL}$, would be approximately 28.6\,MHz. As shown in Figure \ref{RAM plots}, the RAM noise witnessed at the cavity reflection is similar to the noise measured directly after the EOM, and the measured cavity noise is significantly improved in both OTI cavities, when using a modulation frequency of 28.6\,MHz.

\begin{figure}[h]
    \centering
    \includegraphics[width=\linewidth]{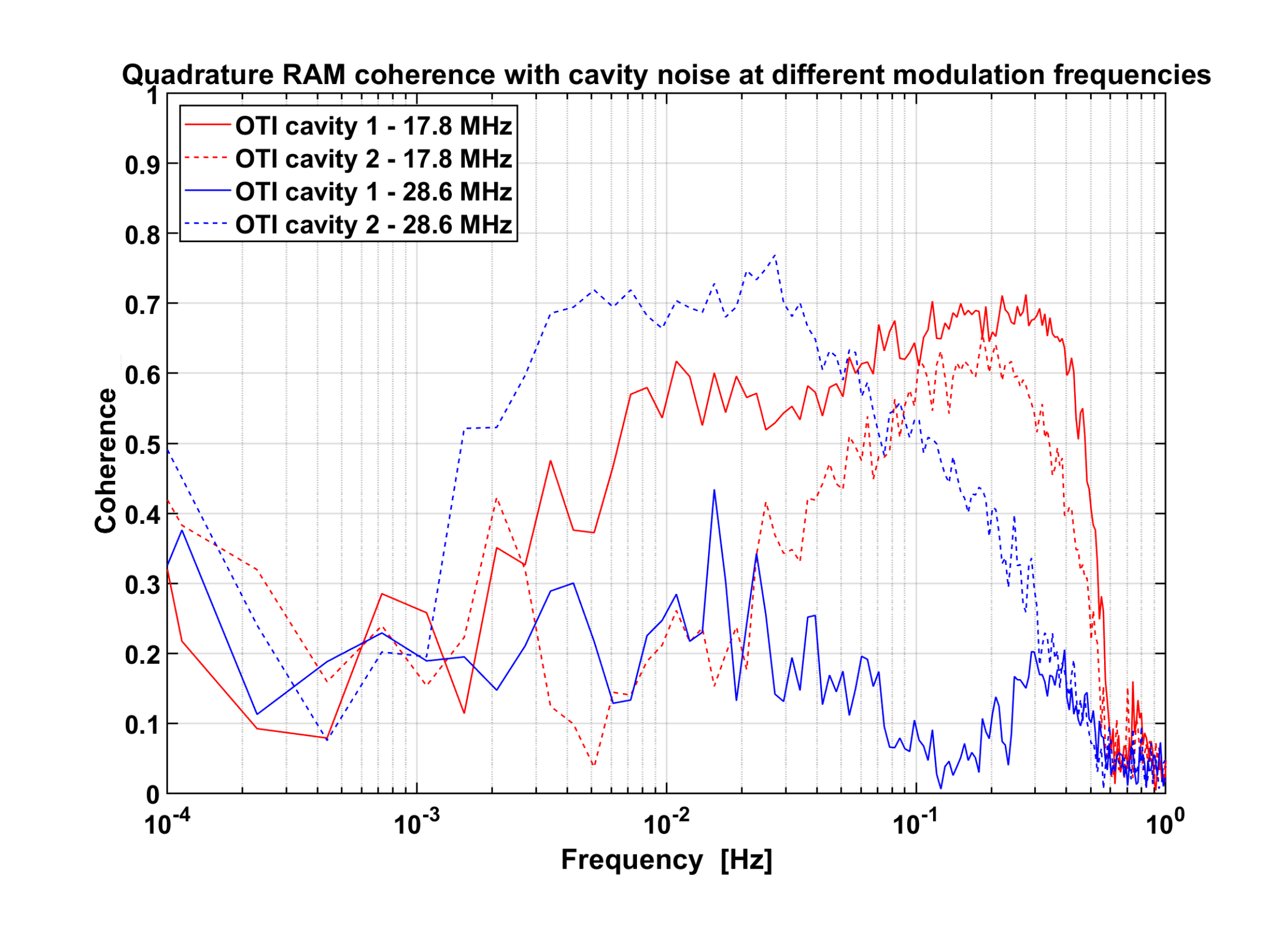}
    \caption{The coherence between the quadrature component of the RAM at the cavity reflection and the cavity noise. The red curves are from measurements with a modulation frequency of 17.8 MHz and the blue curves are with a frequency of 28.6 MHz.}
    \label{RAM coherence}
\end{figure}

We also monitored the quadrature component of the RAM in the cavity reflection concurrently with our cavity beat note measurements to find their coherence in the LISA band. Figure \ref{RAM coherence} shows the coherence estimates between the quadrature RAM and the length noise measured in both OTI cavities and compares the estimates when using the different phase modulation frequencies. Since we cannot measure the in-phase RAM concurrently with the beat note measurements, these coherence estimates do not completely characterize how the RAM contributes to the cavity measurements. However, we can see that when switching from a modulation frequency of 17.8\,MHz to 28.6\,MHz, the RAM coherence with OTI 1 is significantly reduced while the coherence with OTI 2 is larger around 2\,mHz - 0.1\,Hz. This suggests that the in-phase RAM was likely the dominant source of noise for OTI 2 when using the original modulation frequency, and the quadrature component has a higher correlation after reducing the in-phase noise. The difference in noise between OTI 1 and OTI 2 is likely explained by the under-coupling of OTI 2, as that would generally lead to a larger susceptibility to RAM than in an impedance-matched or over-coupled cavity \cite{shi_suppression_2018}. Overall, we have been able to identify and characterize the RAM as a dominant source of noise in our early measurements of the OTI cavities, and we have significantly reduced this noise such that we can measure each cavity well below the $1 \frac{\mathrm{pm}}{\sqrt{\mathrm{Hz}}}$ LISA sensitivity requirement above 1 mHz.

\subsection{Other Noise Sources}
\label{sec:3.4}

Aside from the primary noise sources that are discussed in Sections \ref{sec:3.1}-\ref{sec:3.3}, we also estimated the contributions from the dark noise of the PDs as well as the shot noise in the cavity reflection. The contribution from the dark noise was estimated by blocking any light from hitting the PD and measuring the demodulated signal that would otherwise be used for PDH locking. This is measured as a voltage time series, and the ASD is then scaled by the cavity discriminant that is also used to scale the RAM noise estimation, both of which are shown in Figure \ref{Noise budgets}. The shot noise contribution can be roughly estimated by following Black (2001), and we calculate a contribution on the order of $10^{-18}$ $\frac{\mathrm{m}}{\sqrt{\mathrm{Hz}}}$, far below any other noise estimation we have considered and thus regarded as negligible. One source of noise that was not considered in our investigations was scattered light. The phase-modulated beams can scatter from imperfections in the fiber-coupled components and the mating sleeve links between them, and the imperfections in the cavity mirrors can also cause scattering inside the cavities. Much of this scattered light is propagated away from the primary beam and is absorbed by the surrounding environment, while a small fraction of the scattering may couple back into the primary mode. The PD measuring the cavity reflection will then witness a small fraction of scattered light at the carrier and side-band frequencies that interferes with the primary beam at a spurious phase. This interference can contaminate the PDH signal and affect the measured cavity noise similarly to how the RAM noise couples into the cavity noise. However, a detailed analysis of the contributions from the scattered light to our measurements of the OTI cavities is beyond the scope of this paper and will be left for future investigations.

\section{Conclusions \& Future Work}
\label{sec:conc}

We have developed and tested the performance of two prototype optical truss cavities, both of which have been shown to meet the $1 \frac{\mathrm{pm}}{\sqrt{\mathrm{Hz}}}$ LISA stability requirement at 0.1\,mHz - 1\,Hz. These cavities are constructed with compact, fiber-coupled input stages mounted to plates of ULE glass with simple stainless-steel clamps. The first cavity, OTI 1, was closed with our originally designed return stage which matches the input stage. The second cavity, OTI 2, was closed with an alternative return mirror and mount to account for a shorter cavity baseline than the original design specifications. These two cavities represent a powerful ``plug and play'' solution for measuring picometer stability in the LISA band. The OTI units can be integrated with virtually any stable structure, given a properly designed mounting apparatus and return stage, and only requires interfacing with fiber-coupled components and PDH locking electronics. The primary source of noise above 1\,mHz was found to be the residual amplitude modulation (RAM), which was exacerbated by a parasitic etalon interference in the fiber circulators used. We solved this issue by carefully choosing the phase modulation frequency to match the free spectral range (FSR) of the parasitic etalon, which significantly improved the measured cavity noise by roughly a factor of 2 as shown in Figure \ref{RAM plots}. Our results and analyses highlight the possibilities for the future development of ground testing experiments that can be interfaced with compact, fiber-based ``plug and play'' cavity systems.

In the contingency that the OTI must be integrated with the LISA mission, all three cavities must be read out with one pre-stabilized laser source. In the experiments discussed here, we used one laser per cavity and actuated each laser controller with their respective PDH feedback signals. For LISA, this would require six additional lasers per spacecraft (three OTI cavities per telescope). Future work on the OTI prototypes will seek to test novel offset-locking methods to read out the cavity displacement noise in both OTI 1 and OTI 2 with only one laser source which would be PDH-locked to the reference cavity \cite{edwards_aps_2023,thorpe_laser_2008}. Similarly to the measurements outlined in this paper, we will aim to demonstrate a $1 \frac{\mathrm{pm}}{\sqrt{\mathrm{Hz}}}$ sensitivity in both prototype OTI cavities utilizing a pre-stabilized laser source, thus solidifying a complete proof-of-concept for the optical truss interferometer.

\section*{Funding}
The authors gratefully acknowledge the financial support from the NASA Goddard Space Flight Center (Grant 80NSSC22K0675) and the University of Florida (Sub-Award SUB00003369).

\section*{Acknowledgments}
The authors acknowledge Dr. Soham Kulkarni, Dr. Ada Agnieszka Uminska, Joseph Gleason, Laura Rachelle Roberts, Daniel George, Paul Edwards, Dr. Alexander Weaver, Prof. David Tanner, and Sourath Ghosh for their helpful contributions and advice throughout our experimental work. The authors also acknowledge Bill Malphurs and the team at the UF machine shop for their help in the design and machining of critical parts for the experiment.

\section*{Disclosures}
The authors declare no conflicts of interest.

\section*{Data Availability}
Data underlying the results presented in this paper are not publicly available at this time but may be obtained from the authors upon reasonable request.

\newpage

\bibliography{Bibliography}

\begin{thebibliography}{34}%
\makeatletter
\providecommand \@ifxundefined [1]{%
 \@ifx{#1\undefined}
}%
\providecommand \@ifnum [1]{%
 \ifnum #1\expandafter \@firstoftwo
 \else \expandafter \@secondoftwo
 \fi
}%
\providecommand \@ifx [1]{%
 \ifx #1\expandafter \@firstoftwo
 \else \expandafter \@secondoftwo
 \fi
}%
\providecommand \natexlab [1]{#1}%
\providecommand \enquote  [1]{``#1''}%
\providecommand \bibnamefont  [1]{#1}%
\providecommand \bibfnamefont [1]{#1}%
\providecommand \citenamefont [1]{#1}%
\providecommand \href@noop [0]{\@secondoftwo}%
\providecommand \href [0]{\begingroup \@sanitize@url \@href}%
\providecommand \@href[1]{\@@startlink{#1}\@@href}%
\providecommand \@@href[1]{\endgroup#1\@@endlink}%
\providecommand \@sanitize@url [0]{\catcode `\\12\catcode `\$12\catcode `\&12\catcode `\#12\catcode `\^12\catcode `\_12\catcode `\%12\relax}%
\providecommand \@@startlink[1]{}%
\providecommand \@@endlink[0]{}%
\providecommand \url  [0]{\begingroup\@sanitize@url \@url }%
\providecommand \@url [1]{\endgroup\@href {#1}{\urlprefix }}%
\providecommand \urlprefix  [0]{URL }%
\providecommand \Eprint [0]{\href }%
\providecommand \doibase [0]{http://dx.doi.org/}%
\providecommand \selectlanguage [0]{\@gobble}%
\providecommand \bibinfo  [0]{\@secondoftwo}%
\providecommand \bibfield  [0]{\@secondoftwo}%
\providecommand \translation [1]{[#1]}%
\providecommand \BibitemOpen [0]{}%
\providecommand \bibitemStop [0]{}%
\providecommand \bibitemNoStop [0]{.\EOS\space}%
\providecommand \EOS [0]{\spacefactor3000\relax}%
\providecommand \BibitemShut  [1]{\csname bibitem#1\endcsname}%
\let\auto@bib@innerbib\@empty
%</preamble>
\bibitem [{\citenamefont {Thorpe}(2010)}]{thorpe_lisa_2010}%
  \BibitemOpen
  \bibfield  {author} {\bibinfo {author} {\bibfnamefont {J.~I.}\ \bibnamefont {Thorpe}},\ }\href {\doibase 10.1088/0264-9381/27/8/084008} {\bibfield  {journal} {\bibinfo  {journal} {Class. Quantum Grav.}\ }\textbf {\bibinfo {volume} {27}},\ \bibinfo {pages} {084008} (\bibinfo {year} {2010})}\BibitemShut {NoStop}%
\bibitem [{\citenamefont {Shaddock}(2008)}]{shaddock_space-based_2008}%
  \BibitemOpen
  \bibfield  {author} {\bibinfo {author} {\bibfnamefont {D.~A.}\ \bibnamefont {Shaddock}},\ }\href {\doibase 10.1088/0264-9381/25/11/114012} {\bibfield  {journal} {\bibinfo  {journal} {Class. Quantum Grav.}\ }\textbf {\bibinfo {volume} {25}},\ \bibinfo {pages} {114012} (\bibinfo {year} {2008})}\BibitemShut {NoStop}%
\bibitem [{\citenamefont {Wanner}(2019)}]{wanner_space-based_2019}%
  \BibitemOpen
  \bibfield  {author} {\bibinfo {author} {\bibfnamefont {G.}~\bibnamefont {Wanner}},\ }\href {\doibase 10.1038/s41567-019-0462-3} {\bibfield  {journal} {\bibinfo  {journal} {Nat. Phys.}\ }\textbf {\bibinfo {volume} {15}},\ \bibinfo {pages} {200} (\bibinfo {year} {2019})}\BibitemShut {NoStop}%
\bibitem [{\citenamefont {Jennrich}(2009)}]{jennrich_lisa_2009}%
  \BibitemOpen
  \bibfield  {author} {\bibinfo {author} {\bibfnamefont {O.}~\bibnamefont {Jennrich}},\ }\href {\doibase 10.1088/0264-9381/26/15/153001} {\bibfield  {journal} {\bibinfo  {journal} {Class. Quantum Grav.}\ }\textbf {\bibinfo {volume} {26}},\ \bibinfo {pages} {153001} (\bibinfo {year} {2009})}\BibitemShut {NoStop}%
\bibitem [{\citenamefont {Jersey}\ \emph {et~al.}(2023)\citenamefont {Jersey}, \citenamefont {Harley-Trochimczyk}, \citenamefont {Zhang},\ and\ \citenamefont {Guzman}}]{jersey_optical_2023}%
  \BibitemOpen
  \bibfield  {author} {\bibinfo {author} {\bibfnamefont {K.}~\bibnamefont {Jersey}}, \bibinfo {author} {\bibfnamefont {I.}~\bibnamefont {Harley-Trochimczyk}}, \bibinfo {author} {\bibfnamefont {Y.}~\bibnamefont {Zhang}}, \ and\ \bibinfo {author} {\bibfnamefont {F.}~\bibnamefont {Guzman}},\ }\href {\doibase 10.1364/AO.493108} {\bibfield  {journal} {\bibinfo  {journal} {Appl. Opt.}\ }\textbf {\bibinfo {volume} {62}},\ \bibinfo {pages} {5675} (\bibinfo {year} {2023})}\BibitemShut {NoStop}%
\bibitem [{\citenamefont {Elliffe}\ \emph {et~al.}(2005)\citenamefont {Elliffe}, \citenamefont {Bogenstahl}, \citenamefont {Deshpande}, \citenamefont {Hough}, \citenamefont {Killow}, \citenamefont {Reid}, \citenamefont {Robertson}, \citenamefont {Rowan}, \citenamefont {Ward},\ and\ \citenamefont {Cagnoli}}]{elliffe_hydroxide-catalysis_2005}%
  \BibitemOpen
  \bibfield  {author} {\bibinfo {author} {\bibfnamefont {E.~J.}\ \bibnamefont {Elliffe}}, \bibinfo {author} {\bibfnamefont {J.}~\bibnamefont {Bogenstahl}}, \bibinfo {author} {\bibfnamefont {A.}~\bibnamefont {Deshpande}}, \bibinfo {author} {\bibfnamefont {J.}~\bibnamefont {Hough}}, \bibinfo {author} {\bibfnamefont {C.}~\bibnamefont {Killow}}, \bibinfo {author} {\bibfnamefont {S.}~\bibnamefont {Reid}}, \bibinfo {author} {\bibfnamefont {D.}~\bibnamefont {Robertson}}, \bibinfo {author} {\bibfnamefont {S.}~\bibnamefont {Rowan}}, \bibinfo {author} {\bibfnamefont {H.}~\bibnamefont {Ward}}, \ and\ \bibinfo {author} {\bibfnamefont {G.}~\bibnamefont {Cagnoli}},\ }\href {\doibase 10.1088/0264-9381/22/10/018} {\bibfield  {journal} {\bibinfo  {journal} {Class. Quantum Grav.}\ }\textbf {\bibinfo {volume} {22}},\ \bibinfo {pages} {S257} (\bibinfo {year} {2005})}\BibitemShut {NoStop}%
\bibitem [{\citenamefont {{Schott}}()}]{zerodur}%
  \BibitemOpen
  \bibfield  {author} {\bibinfo {author} {\bibnamefont {{Schott}}},\ }\href@noop {} {\enquote {\bibinfo {title} {{ZERODUR}},}\ }\bibinfo {howpublished} {\url{https://www.schott.com/en-au/products/zerodur-P1000269}}\BibitemShut {NoStop}%
\bibitem [{\citenamefont {{Corning}}({\natexlab{a}})}]{ULE_glass}%
  \BibitemOpen
  \bibfield  {author} {\bibinfo {author} {\bibnamefont {{Corning}}},\ }\href@noop {} {\enquote {\bibinfo {title} {{ULE Glass}},}\ }\bibinfo {howpublished} {\url{https://www.corning.com/worldwide/en/products/advanced-optics/product-materials/semiconductor-laser-optic-components/ultra-low-expansion-glass.html}} ({\natexlab{a}})\BibitemShut {NoStop}%
\bibitem [{\citenamefont {Kulkarni}\ \emph {et~al.}(2020)\citenamefont {Kulkarni}, \citenamefont {Umińska}, \citenamefont {Gleason}, \citenamefont {Barke}, \citenamefont {Ferguson}, \citenamefont {Sanjuán}, \citenamefont {Fulda},\ and\ \citenamefont {Mueller}}]{kulkarni_ultrastable_2020}%
  \BibitemOpen
  \bibfield  {author} {\bibinfo {author} {\bibfnamefont {S.}~\bibnamefont {Kulkarni}}, \bibinfo {author} {\bibfnamefont {A.}~\bibnamefont {Umińska}}, \bibinfo {author} {\bibfnamefont {J.}~\bibnamefont {Gleason}}, \bibinfo {author} {\bibfnamefont {S.}~\bibnamefont {Barke}}, \bibinfo {author} {\bibfnamefont {R.}~\bibnamefont {Ferguson}}, \bibinfo {author} {\bibfnamefont {J.}~\bibnamefont {Sanjuán}}, \bibinfo {author} {\bibfnamefont {P.}~\bibnamefont {Fulda}}, \ and\ \bibinfo {author} {\bibfnamefont {G.}~\bibnamefont {Mueller}},\ }\href {\doibase 10.1364/AO.395831} {\bibfield  {journal} {\bibinfo  {journal} {Appl. Opt.}\ }\textbf {\bibinfo {volume} {59}},\ \bibinfo {pages} {6999} (\bibinfo {year} {2020})}\BibitemShut {NoStop}%
\bibitem [{\citenamefont {Kulkarni}(2022)}]{kulkarni_technology_2022}%
  \BibitemOpen
  \bibfield  {author} {\bibinfo {author} {\bibfnamefont {S.}~\bibnamefont {Kulkarni}},\ }\href@noop {} {\enquote {\bibinfo {title} {{Technology Development for Ground Verification of Dimensional Stability of the LISA Telescope}},}\ } (\bibinfo {year} {2022}),\ \bibinfo {note} {{Ph.D. Thesis, University of Florida}}\BibitemShut {NoStop}%
\bibitem [{\citenamefont {Peabody}\ and\ \citenamefont {Merkowitz}(2006)}]{peabody_low_2006}%
  \BibitemOpen
  \bibfield  {author} {\bibinfo {author} {\bibfnamefont {H.}~\bibnamefont {Peabody}}\ and\ \bibinfo {author} {\bibfnamefont {S.~M.}\ \bibnamefont {Merkowitz}},\ }in\ \href {\doibase 10.1063/1.2405044} {\emph {\bibinfo {booktitle} {{AIP} {Conference} {Proceedings}}}},\ Vol.\ \bibinfo {volume} {873}\ (\bibinfo  {publisher} {AIP},\ \bibinfo {address} {Greenbelt, Maryland (USA)},\ \bibinfo {year} {2006})\ pp.\ \bibinfo {pages} {204--209},\ \bibinfo {note} {iSSN: 0094243X}\BibitemShut {NoStop}%
\bibitem [{\citenamefont {Sanjuan}\ \emph {et~al.}(2015)\citenamefont {Sanjuan}, \citenamefont {Gürlebeck},\ and\ \citenamefont {Braxmaier}}]{sanjuan_mathematical_2015}%
  \BibitemOpen
  \bibfield  {author} {\bibinfo {author} {\bibfnamefont {J.}~\bibnamefont {Sanjuan}}, \bibinfo {author} {\bibfnamefont {N.}~\bibnamefont {Gürlebeck}}, \ and\ \bibinfo {author} {\bibfnamefont {C.}~\bibnamefont {Braxmaier}},\ }\href {\doibase 10.1364/OE.23.017892} {\bibfield  {journal} {\bibinfo  {journal} {Opt. Express}\ }\textbf {\bibinfo {volume} {23}},\ \bibinfo {pages} {17892} (\bibinfo {year} {2015})}\BibitemShut {NoStop}%
\bibitem [{\citenamefont {{Corning}}({\natexlab{b}})}]{macor}%
  \BibitemOpen
  \bibfield  {author} {\bibinfo {author} {\bibnamefont {{Corning}}},\ }\href@noop {} {\enquote {\bibinfo {title} {{MACOR Machinable Glass Ceramic}},}\ }\bibinfo {howpublished} {\url{https://www.corning.com/worldwide/en/products/advanced-optics/product-materials/specialty-glass-and-glass-ceramics/glass-ceramics/macor.html}} ({\natexlab{b}})\BibitemShut {NoStop}%
\bibitem [{\citenamefont {Steinhart}\ and\ \citenamefont {Hart}(1968)}]{steinhart_calibration_1968}%
  \BibitemOpen
  \bibfield  {author} {\bibinfo {author} {\bibfnamefont {J.~S.}\ \bibnamefont {Steinhart}}\ and\ \bibinfo {author} {\bibfnamefont {S.~R.}\ \bibnamefont {Hart}},\ }\href {\doibase 10.1016/0011-7471(68)90057-0} {\bibfield  {journal} {\bibinfo  {journal} {Deep Sea Research and Oceanographic Abstracts}\ }\textbf {\bibinfo {volume} {15}},\ \bibinfo {pages} {497} (\bibinfo {year} {1968})}\BibitemShut {NoStop}%
\bibitem [{\citenamefont {Sanjuan}\ \emph {et~al.}(2019)\citenamefont {Sanjuan}, \citenamefont {Abich}, \citenamefont {Gohlke}, \citenamefont {Resch}, \citenamefont {Schuldt}, \citenamefont {Wegehaupt}, \citenamefont {Barwood}, \citenamefont {Gill},\ and\ \citenamefont {Braxmaier}}]{sanjuan_long-term_2019}%
  \BibitemOpen
  \bibfield  {author} {\bibinfo {author} {\bibfnamefont {J.}~\bibnamefont {Sanjuan}}, \bibinfo {author} {\bibfnamefont {K.}~\bibnamefont {Abich}}, \bibinfo {author} {\bibfnamefont {M.}~\bibnamefont {Gohlke}}, \bibinfo {author} {\bibfnamefont {A.}~\bibnamefont {Resch}}, \bibinfo {author} {\bibfnamefont {T.}~\bibnamefont {Schuldt}}, \bibinfo {author} {\bibfnamefont {T.}~\bibnamefont {Wegehaupt}}, \bibinfo {author} {\bibfnamefont {G.~P.}\ \bibnamefont {Barwood}}, \bibinfo {author} {\bibfnamefont {P.}~\bibnamefont {Gill}}, \ and\ \bibinfo {author} {\bibfnamefont {C.}~\bibnamefont {Braxmaier}},\ }\href {\doibase 10.1364/OE.27.036206} {\bibfield  {journal} {\bibinfo  {journal} {Opt. Express}\ }\textbf {\bibinfo {volume} {27}},\ \bibinfo {pages} {36206} (\bibinfo {year} {2019})}\BibitemShut {NoStop}%
\bibitem [{\citenamefont {Numata}\ \emph {et~al.}(2004)\citenamefont {Numata}, \citenamefont {Kemery},\ and\ \citenamefont {Camp}}]{numata_thermal-noise_2004}%
  \BibitemOpen
  \bibfield  {author} {\bibinfo {author} {\bibfnamefont {K.}~\bibnamefont {Numata}}, \bibinfo {author} {\bibfnamefont {A.}~\bibnamefont {Kemery}}, \ and\ \bibinfo {author} {\bibfnamefont {J.}~\bibnamefont {Camp}},\ }\href {\doibase 10.1103/PhysRevLett.93.250602} {\bibfield  {journal} {\bibinfo  {journal} {Phys. Rev. Lett.}\ }\textbf {\bibinfo {volume} {93}},\ \bibinfo {pages} {250602} (\bibinfo {year} {2004})}\BibitemShut {NoStop}%
\bibitem [{\citenamefont {Kessler}\ \emph {et~al.}(2012)\citenamefont {Kessler}, \citenamefont {Legero},\ and\ \citenamefont {Sterr}}]{kessler_thermal_2012}%
  \BibitemOpen
  \bibfield  {author} {\bibinfo {author} {\bibfnamefont {T.}~\bibnamefont {Kessler}}, \bibinfo {author} {\bibfnamefont {T.}~\bibnamefont {Legero}}, \ and\ \bibinfo {author} {\bibfnamefont {U.}~\bibnamefont {Sterr}},\ }\href {\doibase 10.1364/JOSAB.29.000178} {\bibfield  {journal} {\bibinfo  {journal} {J. Opt. Soc. Am. B}\ }\textbf {\bibinfo {volume} {29}},\ \bibinfo {pages} {178} (\bibinfo {year} {2012})}\BibitemShut {NoStop}%
\bibitem [{\citenamefont {Black}(2001)}]{black_introduction_2001}%
  \BibitemOpen
  \bibfield  {author} {\bibinfo {author} {\bibfnamefont {E.~D.}\ \bibnamefont {Black}},\ }\href {\doibase 10.1119/1.1286663} {\bibfield  {journal} {\bibinfo  {journal} {American Journal of Physics}\ }\textbf {\bibinfo {volume} {69}},\ \bibinfo {pages} {79} (\bibinfo {year} {2001})}\BibitemShut {NoStop}%
\bibitem [{\citenamefont {{Liquid Instruments}}()}]{moku_phasemeter}%
  \BibitemOpen
  \bibfield  {author} {\bibinfo {author} {\bibnamefont {{Liquid Instruments}}},\ }\href@noop {} {\enquote {\bibinfo {title} {{Moku Phasemeter}},}\ }\bibinfo {howpublished} {\url{https://www.liquidinstruments.com/products/integrated-instruments/phasemeter/}}\BibitemShut {NoStop}%
\bibitem [{\citenamefont {Stoica}\ and\ \citenamefont {Moses}(2005)}]{stoica_spectral_2005}%
  \BibitemOpen
  \bibfield  {author} {\bibinfo {author} {\bibfnamefont {P.}~\bibnamefont {Stoica}}\ and\ \bibinfo {author} {\bibfnamefont {R.~L.}\ \bibnamefont {Moses}},\ }\href@noop {} {\emph {\bibinfo {title} {Spectral analysis of signals}}}\ (\bibinfo  {publisher} {Pearson/Prentice Hall},\ \bibinfo {address} {Upper Saddle River, N.J},\ \bibinfo {year} {2005})\BibitemShut {NoStop}%
\bibitem [{\citenamefont {Welch}(1967)}]{welch_use_1967}%
  \BibitemOpen
  \bibfield  {author} {\bibinfo {author} {\bibfnamefont {P.}~\bibnamefont {Welch}},\ }\href {\doibase 10.1109/TAU.1967.1161901} {\bibfield  {journal} {\bibinfo  {journal} {IEEE Trans. Audio Electroacoust.}\ }\textbf {\bibinfo {volume} {15}},\ \bibinfo {pages} {70} (\bibinfo {year} {1967})}\BibitemShut {NoStop}%
\bibitem [{\citenamefont {Hewitson}\ \emph {et~al.}(2009)\citenamefont {Hewitson} \emph {et~al.}}]{hewitson_data_2009}%
  \BibitemOpen
  \bibfield  {author} {\bibinfo {author} {\bibfnamefont {M.}~\bibnamefont {Hewitson}} \emph {et~al.},\ }\href {\doibase 10.1088/0264-9381/26/9/094003} {\bibfield  {journal} {\bibinfo  {journal} {Class. Quantum Grav.}\ }\textbf {\bibinfo {volume} {26}},\ \bibinfo {pages} {094003} (\bibinfo {year} {2009})}\BibitemShut {NoStop}%
\bibitem [{\citenamefont {Gretarsson}\ and\ \citenamefont {Gretarsson}(2018)}]{gretarsson_three_2018}%
  \BibitemOpen
  \bibfield  {author} {\bibinfo {author} {\bibfnamefont {E.~M.}\ \bibnamefont {Gretarsson}}\ and\ \bibinfo {author} {\bibfnamefont {A.~M.}\ \bibnamefont {Gretarsson}},\ }\href {\doibase 10.1103/PhysRevD.98.122004} {\bibfield  {journal} {\bibinfo  {journal} {Phys. Rev. D}\ }\textbf {\bibinfo {volume} {98}},\ \bibinfo {pages} {122004} (\bibinfo {year} {2018})}\BibitemShut {NoStop}%
\bibitem [{\citenamefont {Evans}\ \emph {et~al.}(2008)\citenamefont {Evans}, \citenamefont {Ballmer}, \citenamefont {Fejer}, \citenamefont {Fritschel}, \citenamefont {Harry},\ and\ \citenamefont {Ogin}}]{evans_thermo-optic_2008}%
  \BibitemOpen
  \bibfield  {author} {\bibinfo {author} {\bibfnamefont {M.}~\bibnamefont {Evans}}, \bibinfo {author} {\bibfnamefont {S.}~\bibnamefont {Ballmer}}, \bibinfo {author} {\bibfnamefont {M.}~\bibnamefont {Fejer}}, \bibinfo {author} {\bibfnamefont {P.}~\bibnamefont {Fritschel}}, \bibinfo {author} {\bibfnamefont {G.}~\bibnamefont {Harry}}, \ and\ \bibinfo {author} {\bibfnamefont {G.}~\bibnamefont {Ogin}},\ }\href {\doibase 10.1103/PhysRevD.78.102003} {\bibfield  {journal} {\bibinfo  {journal} {Phys. Rev. D}\ }\textbf {\bibinfo {volume} {78}},\ \bibinfo {pages} {102003} (\bibinfo {year} {2008})}\BibitemShut {NoStop}%
\bibitem [{\citenamefont {Hello}\ and\ \citenamefont {Vinet}(1990)}]{hello_analytical_1990}%
  \BibitemOpen
  \bibfield  {author} {\bibinfo {author} {\bibfnamefont {P.}~\bibnamefont {Hello}}\ and\ \bibinfo {author} {\bibfnamefont {J.-Y.}\ \bibnamefont {Vinet}},\ }\href {\doibase 10.1051/jphys:0199000510200224300} {\bibfield  {journal} {\bibinfo  {journal} {J. Phys. France}\ }\textbf {\bibinfo {volume} {51}},\ \bibinfo {pages} {2243} (\bibinfo {year} {1990})}\BibitemShut {NoStop}%
\bibitem [{\citenamefont {Zhang}\ \emph {et~al.}(2014)\citenamefont {Zhang}, \citenamefont {Martin}, \citenamefont {Benko}, \citenamefont {Hall}, \citenamefont {Ye}, \citenamefont {Hagemann}, \citenamefont {Legero}, \citenamefont {Sterr}, \citenamefont {Riehle}, \citenamefont {Cole},\ and\ \citenamefont {Aspelmeyer}}]{zhang_reduction_2014}%
  \BibitemOpen
  \bibfield  {author} {\bibinfo {author} {\bibfnamefont {W.}~\bibnamefont {Zhang}}, \bibinfo {author} {\bibfnamefont {M.~J.}\ \bibnamefont {Martin}}, \bibinfo {author} {\bibfnamefont {C.}~\bibnamefont {Benko}}, \bibinfo {author} {\bibfnamefont {J.~L.}\ \bibnamefont {Hall}}, \bibinfo {author} {\bibfnamefont {J.}~\bibnamefont {Ye}}, \bibinfo {author} {\bibfnamefont {C.}~\bibnamefont {Hagemann}}, \bibinfo {author} {\bibfnamefont {T.}~\bibnamefont {Legero}}, \bibinfo {author} {\bibfnamefont {U.}~\bibnamefont {Sterr}}, \bibinfo {author} {\bibfnamefont {F.}~\bibnamefont {Riehle}}, \bibinfo {author} {\bibfnamefont {G.~D.}\ \bibnamefont {Cole}}, \ and\ \bibinfo {author} {\bibfnamefont {M.}~\bibnamefont {Aspelmeyer}},\ }\href {\doibase 10.1364/OL.39.001980} {\bibfield  {journal} {\bibinfo  {journal} {Opt. Lett.}\ }\textbf {\bibinfo {volume} {39}},\ \bibinfo {pages} {1980} (\bibinfo {year} {2014})}\BibitemShut {NoStop}%
\bibitem [{\citenamefont {Wong}\ and\ \citenamefont {Hall}(1985)}]{wong_servo_1985}%
  \BibitemOpen
  \bibfield  {author} {\bibinfo {author} {\bibfnamefont {N.~C.}\ \bibnamefont {Wong}}\ and\ \bibinfo {author} {\bibfnamefont {J.~L.}\ \bibnamefont {Hall}},\ }\href {\doibase 10.1364/JOSAB.2.001527} {\bibfield  {journal} {\bibinfo  {journal} {J. Opt. Soc. Am. B, JOSAB}\ }\textbf {\bibinfo {volume} {2}},\ \bibinfo {pages} {1527} (\bibinfo {year} {1985})}\BibitemShut {NoStop}%
\bibitem [{\citenamefont {Duong}\ \emph {et~al.}(2018)\citenamefont {Duong}, \citenamefont {Nguyen}, \citenamefont {Vu}, \citenamefont {Higuchi}, \citenamefont {Wei},\ and\ \citenamefont {Aketagawa}}]{duong_suppression_2018}%
  \BibitemOpen
  \bibfield  {author} {\bibinfo {author} {\bibfnamefont {Q.~A.}\ \bibnamefont {Duong}}, \bibinfo {author} {\bibfnamefont {T.~D.}\ \bibnamefont {Nguyen}}, \bibinfo {author} {\bibfnamefont {T.~T.}\ \bibnamefont {Vu}}, \bibinfo {author} {\bibfnamefont {M.}~\bibnamefont {Higuchi}}, \bibinfo {author} {\bibfnamefont {D.}~\bibnamefont {Wei}}, \ and\ \bibinfo {author} {\bibfnamefont {M.}~\bibnamefont {Aketagawa}},\ }\href {\doibase 10.1186/s41476-018-0092-x} {\bibfield  {journal} {\bibinfo  {journal} {J. Eur. Opt. Soc.-Rapid Publ.}\ }\textbf {\bibinfo {volume} {14}},\ \bibinfo {pages} {25} (\bibinfo {year} {2018})}\BibitemShut {NoStop}%
\bibitem [{\citenamefont {Shi}\ \emph {et~al.}(2018)\citenamefont {Shi}, \citenamefont {Zhang}, \citenamefont {Zeng}, \citenamefont {Lü}, \citenamefont {Liu}, \citenamefont {Xi}, \citenamefont {Ye},\ and\ \citenamefont {Lu}}]{shi_suppression_2018}%
  \BibitemOpen
  \bibfield  {author} {\bibinfo {author} {\bibfnamefont {X.}~\bibnamefont {Shi}}, \bibinfo {author} {\bibfnamefont {J.}~\bibnamefont {Zhang}}, \bibinfo {author} {\bibfnamefont {X.}~\bibnamefont {Zeng}}, \bibinfo {author} {\bibfnamefont {X.}~\bibnamefont {Lü}}, \bibinfo {author} {\bibfnamefont {K.}~\bibnamefont {Liu}}, \bibinfo {author} {\bibfnamefont {J.}~\bibnamefont {Xi}}, \bibinfo {author} {\bibfnamefont {Y.}~\bibnamefont {Ye}}, \ and\ \bibinfo {author} {\bibfnamefont {Z.}~\bibnamefont {Lu}},\ }\href {\doibase 10.1007/s00340-018-7021-y} {\bibfield  {journal} {\bibinfo  {journal} {Appl. Phys. B}\ }\textbf {\bibinfo {volume} {124}},\ \bibinfo {pages} {153} (\bibinfo {year} {2018})}\BibitemShut {NoStop}%
\bibitem [{\citenamefont {Shen}\ \emph {et~al.}(2015)\citenamefont {Shen}, \citenamefont {Li}, \citenamefont {Bi}, \citenamefont {Wang},\ and\ \citenamefont {Chen}}]{shen_systematic_2015}%
  \BibitemOpen
  \bibfield  {author} {\bibinfo {author} {\bibfnamefont {H.}~\bibnamefont {Shen}}, \bibinfo {author} {\bibfnamefont {L.}~\bibnamefont {Li}}, \bibinfo {author} {\bibfnamefont {J.}~\bibnamefont {Bi}}, \bibinfo {author} {\bibfnamefont {J.}~\bibnamefont {Wang}}, \ and\ \bibinfo {author} {\bibfnamefont {L.}~\bibnamefont {Chen}},\ }\href {\doibase 10.1103/PhysRevA.92.063809} {\bibfield  {journal} {\bibinfo  {journal} {Phys. Rev. A}\ }\textbf {\bibinfo {volume} {92}},\ \bibinfo {pages} {063809} (\bibinfo {year} {2015})}\BibitemShut {NoStop}%
\bibitem [{\citenamefont {Wegehaupt}\ \emph {et~al.}(2024)\citenamefont {Wegehaupt}, \citenamefont {Sanjuan}, \citenamefont {Gohlke}, \citenamefont {Grafe}, \citenamefont {Kumanchik}, \citenamefont {Oswald}, \citenamefont {Schuldt},\ and\ \citenamefont {Braxmaier}}]{wegehaupt_optical_2024}%
  \BibitemOpen
  \bibfield  {author} {\bibinfo {author} {\bibfnamefont {T.}~\bibnamefont {Wegehaupt}}, \bibinfo {author} {\bibfnamefont {J.}~\bibnamefont {Sanjuan}}, \bibinfo {author} {\bibfnamefont {M.}~\bibnamefont {Gohlke}}, \bibinfo {author} {\bibfnamefont {P.}~\bibnamefont {Grafe}}, \bibinfo {author} {\bibfnamefont {L.}~\bibnamefont {Kumanchik}}, \bibinfo {author} {\bibfnamefont {M.}~\bibnamefont {Oswald}}, \bibinfo {author} {\bibfnamefont {T.}~\bibnamefont {Schuldt}}, \ and\ \bibinfo {author} {\bibfnamefont {C.}~\bibnamefont {Braxmaier}},\ }\href {\doibase 10.1364/AO.522293} {\bibfield  {journal} {\bibinfo  {journal} {Appl. Opt.}\ }\textbf {\bibinfo {volume} {63}},\ \bibinfo {pages} {3438} (\bibinfo {year} {2024})}\BibitemShut {NoStop}%
\bibitem [{\citenamefont {Whittaker}\ \emph {et~al.}(1985)\citenamefont {Whittaker}, \citenamefont {Gehrtz},\ and\ \citenamefont {Bjorklund}}]{whittaker_residual_1985}%
  \BibitemOpen
  \bibfield  {author} {\bibinfo {author} {\bibfnamefont {E.~A.}\ \bibnamefont {Whittaker}}, \bibinfo {author} {\bibfnamefont {M.}~\bibnamefont {Gehrtz}}, \ and\ \bibinfo {author} {\bibfnamefont {G.~C.}\ \bibnamefont {Bjorklund}},\ }\href {\doibase 10.1364/JOSAB.2.001320} {\bibfield  {journal} {\bibinfo  {journal} {J. Opt. Soc. Am. B}\ }\textbf {\bibinfo {volume} {2}},\ \bibinfo {pages} {1320} (\bibinfo {year} {1985})}\BibitemShut {NoStop}%
\bibitem [{\citenamefont {Edwards}\ and\ \citenamefont {Fulda}(2023)}]{edwards_aps_2023}%
  \BibitemOpen
  \bibfield  {author} {\bibinfo {author} {\bibfnamefont {P.}~\bibnamefont {Edwards}}\ and\ \bibinfo {author} {\bibfnamefont {P.}~\bibnamefont {Fulda}},\ }in\ \href {https://meetings.aps.org/Meeting/APR23/Session/B09.6} {\emph {\bibinfo {booktitle} {APS April Meeting 2023}}},\ Vol.\ \bibinfo {volume} {68, Num. 6}\ (\bibinfo  {publisher} {American Physical Society},\ \bibinfo {year} {2023})\BibitemShut {NoStop}%
\bibitem [{\citenamefont {Thorpe}\ \emph {et~al.}(2008)\citenamefont {Thorpe}, \citenamefont {Numata},\ and\ \citenamefont {Livas}}]{thorpe_laser_2008}%
  \BibitemOpen
  \bibfield  {author} {\bibinfo {author} {\bibfnamefont {J.~I.}\ \bibnamefont {Thorpe}}, \bibinfo {author} {\bibfnamefont {K.}~\bibnamefont {Numata}}, \ and\ \bibinfo {author} {\bibfnamefont {J.}~\bibnamefont {Livas}},\ }\href {\doibase 10.1364/OE.16.015980} {\bibfield  {journal} {\bibinfo  {journal} {Opt. Express}\ }\textbf {\bibinfo {volume} {16}},\ \bibinfo {pages} {15980} (\bibinfo {year} {2008})}\BibitemShut {NoStop}%
\end{thebibliography}%

\end{document}